\documentclass[aps,twocolumn,preprintnumbers,amsmath,amssymb,prb,superscriptaddress]{revtex4}
\usepackage[dvips]{epsfig}
\usepackage{subfigure}
\usepackage{graphicx}
\usepackage{longtable}
\usepackage[ansinew]{inputenc}
\usepackage{ulem}
\usepackage{color}
\usepackage[final]{pdfpages}		

\begin{document}
\title{Exchange Splitting of a Hybrid Surface State and Ferromagnetic Order in a 2D Surface Alloy}

\author{Johannes Knippertz}
\email[]{jseidel@rhrk.uni-kl.de}
\affiliation{Department of Physics and Research Center OPTIMAS, University of Kaiserslautern, Erwin-Schroedinger-Strasse 46, 67663 Kaiserslautern, Germany}
\author{Patrick M. Buhl}
\affiliation{Johannes-Gutenberg-Universit\"at Mainz, Institut f\"ur Physik, 55128 Mainz, Germany}
\author{Sina Mousavion}
\affiliation{Department of Physics and Research Center OPTIMAS, University of Kaiserslautern, Erwin-Schroedinger-Strasse 46, 67663 Kaiserslautern, Germany}
\author{Bertrand Dup\'{e}}
\affiliation{Nanomat/Q-mat/CESAM,Universit\'e de Li\`ege,B-4000 Sart Tilman, Belgium}
\author{Eva S. Walther}
\affiliation{Department of Physics and Research Center OPTIMAS, University of Kaiserslautern, Erwin-Schroedinger-Strasse 46, 67663 Kaiserslautern, Germany}
\author{Katerina Medjanik}
\affiliation{Johannes-Gutenberg-Universit\"at Mainz, Institut f\"ur Physik, 55128 Mainz, Germany}
\author{Dmitry Vasilyev}
\affiliation{Johannes-Gutenberg-Universit\"at Mainz, Institut f\"ur Physik, 55128 Mainz, Germany}
\author{Sergey Babenkov}
\affiliation{Johannes-Gutenberg-Universit\"at Mainz, Institut f\"ur Physik, 55128 Mainz, Germany}
\author{Martin Ellguth}
\affiliation{Johannes-Gutenberg-Universit\"at Mainz, Institut f\"ur Physik, 55128 Mainz, Germany}
\author{Mahalingam Maniraj}
\affiliation{Department of Physics and Research Center OPTIMAS, University of Kaiserslautern, Erwin-Schroedinger-Strasse 46, 67663 Kaiserslautern, Germany}
\author{Jairo Sinova}
\affiliation{Johannes-Gutenberg-Universit\"at Mainz, Institut f\"ur Physik, 55128 Mainz, Germany}
\author{Gerd Sch\"onhense}
\affiliation{Johannes-Gutenberg-Universit\"at Mainz, Institut f\"ur Physik, 55128 Mainz, Germany}
\affiliation{Graduate School of Excellence Materials Science in Mainz}
\author{Hans-Joachim Elmers}
\affiliation{Johannes-Gutenberg-Universit\"at Mainz, Institut f\"ur Physik, 55128 Mainz, Germany}
\affiliation{Graduate School of Excellence Materials Science in Mainz}
\author{Martin Aeschlimann}
\affiliation{Department of Physics and Research Center OPTIMAS, University of Kaiserslautern, Erwin-Schroedinger-Strasse 46, 67663 Kaiserslautern, Germany}
\author{Benjamin Stadtm\"uller}
\email[]{bstadtmueller@physik.uni-kl.de}
\affiliation{Department of Physics and Research Center OPTIMAS, University of Kaiserslautern, Erwin-Schroedinger-Strasse 46, 67663 Kaiserslautern, Germany}
\affiliation{Graduate School of Excellence Materials Science in Mainz}

\date{\today}

\begin{abstract}
Surface alloys are highly flexible materials for tailoring the spin-dependent properties of surfaces. Here, we study the spin-dependent band structure of a DyAg$_2$ surface alloy formed on an Ag(111) crystal. We find a significant exchange spin-splitting of the localized Dy 4f states pointing to a ferromagnetic coupling between the localized Dy moments at $40\,$K. The magnetic coupling between these moments is mediated by an indirect, RKKY-like exchange coupling via the spin-polarized electrons of the hole-like Dy-Ag hybrid surface state.
\end{abstract}

\maketitle

The growing demand for next generation information technology with higher data processing speed and data storage capacity has triggered the quest to design novel low dimensional materials with exotic spin structures to generate and manipulate spin-polarized charge carriers on ever smaller length scales \citep{Wang.2013,Dlubak.2012, Dankert.2017}. In the last decade, surface alloys consisting of heavy metal and noble metal atoms have emerged as a highly tunable class of materials, which have already demonstrated their applicability as nanoscale functional units for spin to charge conversion processes \cite{Sanchez.2013, Manchon.2015}. For these binary materials consisting typically of two host atoms and one alloy atom per surface unit cell, the direct hybridization between the atomic species of the alloy and host materials results in the formation of a hole-like hybrid surface state showing a Rashba-type spin splitting\citep{Ast.2007,Bentmann.2009,Moreschini.2009,Meier.2011,Bihlmayer.2007, Maniraj.2018}. Crucially, the magnitude of this Rashba-type spin splitting depends on the atomic spin-orbit coupling strength\citep{Petersen.2000,Gierz.2011} as well as on the vertical relaxation \citep{ Bentmann.2009, Gierz.2010} of the surface atoms. Both material parameters can be manipulated either by replacing the atomic species of the alloy or the host material or by the formation of tailored bonds between the alloy atoms and molecular adsorbates \citep{Stadtmuller.2016, Friedrich.2017}. These different external control mechanisms hence provide a clear route to functionalize 2D Rashba surface alloys according to the desired field of application. 

So far, the concept of band structure engineering by surface alloying has mainly been limited to non-magnetic systems and has not been explored for ferromagnetic surface alloys. This is mainly due to the fact that the existence of long-range magnetic order in surface alloys was only recently demonstrated for alloys consisting of the rare-earth element gadolinium (Gd) and the noble metals Ag and Au \cite{Ormaza.2016, Correa.2016}. While Ormaza et al. \cite{Ormaza.2016} were able to reveal the Curie temperature as well as the direction of the magnetic anisotropy axis in both systems, the spin-dependent electronic band structure and the microscopic mechanisms leading to ferromagnetic order in low dimensional surface alloys are still elusive. 

In this Letter, we provide a comprehensive view onto the spin-dependent electronic properties of a 2D surface alloy consisting of the rare-earth dysprosium (Dy) and the noble metal silver grown on an Ag(111) surface. The band structure of the 2D surface alloy is investigated by spin- and momentum-resolved photoelectron spectroscopy using the combination of state-of-the-art time-of-flight momentum microscopy \citep{Medjanik.2017,Schonhense.2017} with an imaging spin filter\citep{Kolbe.2011,Kutnyakhov.2013}. This novel experimental approach allows us to access the spin-dependent band structure of the 2D surface alloy throughout the entire Brillouin zone in a fixed experimental geometry. This enables us to disentangle the spectroscopic fingerprints of localized Dy 4f states and of Dy-Ag hybrid surface states from the manifold of back-folded substrate bands. Our experimental findings are discussed in the light of density functional theory (DFT) calculations allowing us to gain insights into the magnetic properties and the reduced electronic correlations of the localized 4f electrons of the 2D surface alloy.

\begin{figure}[ht]
	\centering
		\includegraphics[width=8.5cm]{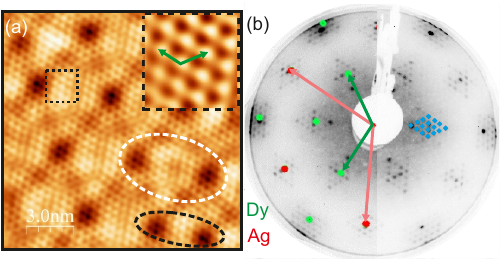}
	\caption{(a) STM images of a DyAg$_2$/Ag(111) film recorded with U$_{\mathrm{Bias}}=30\,$mV (I$=0.07\,$nA). The inset was extracted from the region marked by a black dotted square and shows the local arrangement of the Dy atoms, the dark areas arise due to the moir\'{e} pattern. Black and white ellipsoids highlight the inhomogenous moir\'{e} pattern. (b): LEED pattern of the DyAg$_2$ structure recorded at E=$55\,$eV and room temperature. The left half of the diffraction data is superimposed with the simulated LEED pattern of a $(\sqrt{3}\times\sqrt{3})R30^\circ$ superstructure. The red arrows mark the reciprocal unit cell of Ag(111), the green arrows the one of the DyAg$_2$ surface alloy. The blue dots correspond to a selected number of diffraction spots simulated for a $16\times 16$ superstructure.} 
	\label{fig:Fig1}
\end{figure}

We start our discussion with the structural characterization of the Dy-Ag surface alloy. The adsorption of 1/3 monolayer (ML) of Dy on the clean Ag(111) surface at elevated sample temperature ($575\,$K) results in the formation of the 2D surface alloy with high structural quality and low defect density. A scanning tunneling microscopy (STM) image of the Dy-Ag film is shown in Fig.~1(a). It is dominated by alternating areas with bright and dark contrast which can be attributed to the formation of a long-range ordered moir\'{e} pattern, similar to the one reported recently for other rare-earth noble metal surface alloys\citep{Ormaza.2016}. The long-range order of the moir\'{e} pattern is supported by the well-defined low energy electron diffraction (LEED) pattern in Fig.~1(b). The positions of all diffraction maxima can be explained by a $16\times 16$ superstructure. The simulated diffraction spots are superimposed onto the experimental data in Fig.~1(b) as blue dots. This large unit cell ($A=46.24\,$\AA\, $B=46.24\,$\AA) directly reflects the periodicity of the  moir\'{e} pattern. Most interestingly, the long range order of the moir\'{e} pattern is not affected by local inhomogeneities leading to a non-equidistant distribution between the dark and bright areas, see black and white ellipse in Fig. 1(a). We propose that the complex structure of the moir\'{e} pattern is caused by a lattice mismatch between the Dy-Ag surface alloy and the Ag(111) surface grid. 
The atomic structure itself can be observed in the high-resolution STM image in the inset of Fig.~1(a). It shows a hexagonal lattice of bright protrusions with a periodicity of $(5.1 \pm 0.15)\,$\AA\ (unit cell base vectors indicated in green). The structural parameters of the Dy-Ag superstructure are fully in line with those of a $(\sqrt{3}\times\sqrt{3})R30^\circ$ superstructure, i.e, with the superstructure commonly observed for surface alloys \cite{ Ast.2007,Gierz.2010, Gierz.2011}. Consequently, we propose that the unit cell of the Dy-Ag surface alloy also consists of one Dy atom and two Ag atoms forming a long-range ordered DyAg$_2$ surface alloy on the Ag(111) substrate. 
\begin{figure}[ht]
	\centering
		\includegraphics[width=8.5cm]{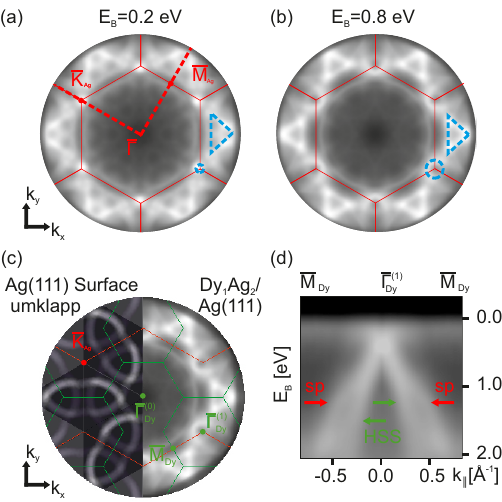}
	\caption{Constant energy maps of the DyAg$_2$ band structure at E$_B=0.2\,$eV (a) and E$_B=0.8\,$eV (b). The surface Brillouin zone of the Ag(111) surface is sketched in red, the high symmetry directions are marked by dashed red lines.(c) Simulation of the surface umklapp processes of the Ag(111) sp-bands due to the formation of a $(\sqrt{3}\times\sqrt{3})R30^\circ$ superstructure. The left half of the CE map shows the measured pattern of the bare Ag(111) bands with additional umklapp structure, the right half the emission pattern of the DyAg$_2$ surface alloy recorded at E$_B=0.8\,$eV. The green hexagons illustrate the SBZs of the DyAg surface alloy. (d) Energy vs. momentum cut along the $\bar{\Gamma}$~$\bar{\mathrm{M}}$ direction of the Dy-Ag Brillouin zone. The green curve indicates the band dispersion of the Dy-Ag hybrid surface state, the red curve the one of a back-folded Ag sp-band. }
	\label{fig:Fig2}
\end{figure}

We now turn to the electronic properties of the DyAg$_2$ surface alloy. The photoemission data were obtained using synchrotron radiation from the BESSY II light source (Helmholtz Center Berlin) with in-plane light polarization (s-polarization) and a photon energy of $\hbar\omega=25\,$eV at a sample temperature of $\approx 40\,$K. The momentum resolved photoemission yield was recorded using a time-of-flight momentum microscope. Exemplary symmetrized constant energy (CE) maps of the momentum microscopy data set are shown in Figs.~2(a) and (b) for two characteristic binding energies of the Dy-Ag surface band structure (a detailed description of the data treatment procedure can be found in the supporting information). The surface Brillouin zone (SBZ) of the Ag(111) surface is marked by a red hexagon and the high symmetry directions $\bar{\Gamma}$~$\bar{\mathrm{M}}_{\text{Ag}}$ and $\overline{\Gamma}$~$\bar{\mathrm{K}}_{\text{Ag}}$ by red dashed lines. Both CE maps reveal an almost vanishing intensity in the center of the SBZ (close to $\bar{\Gamma}$-point) for all binding energies which is attributed to the experimental geometry and the light polarization used in the experiment\cite{ SimonMoser.2017}. The photoemission yield increases significantly for larger momenta, i.e., for the second Brillouin zones, and reveals distinct momentum dependent emission patterns. We find a triangular shaped emission feature for both binding energies along the $\bar{\Gamma}$~$\bar{\mathrm{M}}_{\text{Ag}}$ direction (indicated by a cyan triangle), while the emission feature at the $\bar{\mathrm{K}}_{\text{Ag}}$ point exhibits a ring-like shape at E$_{\mathrm{B}}=0.8\,$eV which transforms into a bright dot at E$_{\mathrm{B}}=0.2\,$eV (indicated by a cyan circle). This energy and momentum dependent intensity at the $\bar{\mathrm{K}}_{\text{Ag}}$-point is consistent with a parabolic dispersion of a hole-like state as expected for a hybrid surface state of a heavy metal/rare-earth alloy formed on noble metal surfaces\citep{Ast.2007,Bentmann.2009,Moreschini.2009,Meier.2011,Bihlmayer.2007, Maniraj.2018,Ormaza.2016}. 

To further support the assignment of this state to an intrinsic feature of the DyAg$_2$ surface alloy, we simulate the expected momentum resolved photoemission yield of the Ag(111) surface due to photoelectron diffraction from Ag bulk states at the periodic Dy-Ag superstructure. This so called surface umklapp process is frequently observed for superstructures on metallic substrates and can severely influence the interpretation of photoemission results of hybrid interfaces \citep{Anderson.1976,Unal.2012,Giovanelli.2013,Stockl.2018}. Our simulation of the surface umklapp process (details in the supporting information) is based on experimental data of the bare Ag(111) surface recorded for identical experimental conditions and is shown in the left half of the CE map in Fig.~2(c). The right half shows the experimental CE map of the DyAg$_2$ surface alloy which we recorded at the same energy (E$_{\mathrm{B}}=0.8\,$eV). Most importantly, no photoemission intensity can be observed at the $\bar{\mathrm{K}}_{\text{Ag}}$-point in the surface umklapp simulations, which proves that the ring-like emission feature at the $\bar{\mathrm{K}}_{\text{Ag}}$-point is an intrinsic spectroscopic signal of the Dy-Ag surface alloy. Crucially, the $\bar{\mathrm{K}}_{\text{Ag}}$-point of the Ag SBZ is equivalent to the $\bar{\Gamma}_{\text{Dy}}^{\left(1\right)}$ of the second SBZ of the DyAg$_2$ surface alloy, i.e., the Dy-Ag hybrid surface state is centered at the $\bar{\Gamma}_{\text{Dy}}^{\left(1\right)}$ point of the SBZ. 

The band dispersion of this hybrid surface state is shown in Fig.~2(d). It clearly reveals the dispersion of a hole-like state (HSS, marked by green arrows), in analogy to all hybrid surface states reported for surface alloys on fcc(111) noble metal surfaces \citep{Ast.2007,Bentmann.2009,Moreschini.2009,Meier.2011,Bihlmayer.2007, Maniraj.2018,Ormaza.2016}. The second band (sp, marked by red arrows) on both sides of the Dy-Ag hybrid state can be identified as a sp-band of the Ag(111) substrate that is back folded at the Dy-Ag superstructure. In addition, we observe a non-dispersing state close to E$_\mathrm{F}$ which we attribute to the spectroscopic signature of the Dy 4f minority state. Interestingly, we only find one non-dispersing Dy 4f state in the energy vs. momentum cut at the $\bar{\Gamma^{\left(1\right)}}$-point of the Dy-Ag surface alloy, and not a corresponding second state as commonly observed for rare earth elements. This can be explained by the characteristic momentum space signature of the second Dy 4f state located at E$_\mathrm{B}=1.8\,$eV, which reveals a significantly smaller photoemission yield at the $\bar{\Gamma}^{\left(1\right)}$-point of the Dy-Ag surface alloy (see additional momentum microscopy data in the supplementary information).

\begin{figure*}[ht]
	\centering
		\includegraphics[width=17cm]{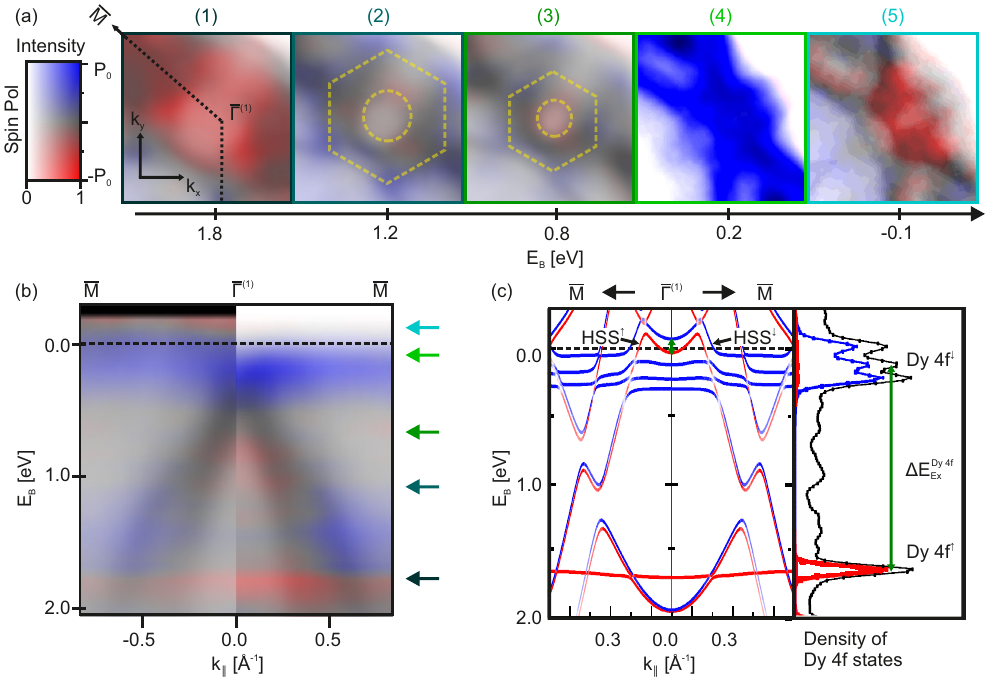}
	\caption{(a) Spin resolved CE maps in the momentum space region of the Dy-Ag hybrid surface state at the $\bar{\Gamma}^{\left(1\right)}$-point. Note that the top most constant energy map has been strongly increased in contrast to compensate the decreased intensity near the Fermi level.  (b) Band structure cut through the hybrid surface state along the $\bar{\Gamma}^{\left(1\right)}$~$\bar{\mathrm{M}}$ direction. Blue color indicates the spin polarization of minority electrons, red contrast the spin polarization of majority electrons. Left half is normalized to enhance features above E$_F$. Arrows on the right indicate the energies of the constant energy maps shown in (a). (c) Spin resolved band structure calculation along the $\bar{\Gamma}^{\left(1\right)}$~$\bar{\mathrm{M}}$ direction (left half) and spin-resolved density of Dy 4f states of the Dy-Ag surface alloy (right half). The colored arrows indicate the binding energy of the CE maps shown in panel (a).}
	\label{fig:Fig3}
\end{figure*}

The spin structure of the Dy-Ag hybrid surface state can be revealed by using the spin-filter branch of the momentum microscope \cite{Schonhense.2017}. Spin selectivity is obtained by the spin-dependent scattering of the photoelectrons at the Ir(001) scattering target for two characteristic scattering energies with distinct spin asymmetry at $12.5\,$eV and a negligible spin asymmetry at $19.0\,$eV \citep{Kutnyakhov.2013} (more experimental details can be found in the supporting information). In this way, we can obtain spin resolved CE maps of the entire valence band structure of the Dy-Ag surface alloy. 

Five selected spin-resolved CE maps in the energy and momentum space region of the Dy-Ag hybrid surface state are shown in Fig.~3(a). The red and blue colored regions in these CE maps indicate bands of positive and negative spin polarization, while grey color corresponds to unpolarized spectral intensity. The CE maps at E$_\mathrm{B}=1.8\,$eV and E$_\mathrm{B}=0.2\,$eV exhibit an almost homogeneous spin contrast of opposite sign, which is not only visible at the positions of the distinct spectroscopic features but also in the photoemission background. This points to the existence of two weakly dispersing states with opposite spin polarization. This conclusion is further supported by the corresponding energy vs. momentum cut in Fig.~3(b). It reveals two non-dispersing states of different spin polarization in narrow energy ranges centered at E$_\mathrm{B}=1.8\,$eV and E$_\mathrm{B}=0.2\,$eV, respectively, which can be attributed to the localized Dy 4f-derived states. The existence of two exchange split Dy 4f states ($\Delta$E$_{\mathrm{Ex}}=1.6\,$eV) with opposite spin polarization in zero external field is a clear spectroscopic signature of ferromagnetic order of the Dy 4f magnetic moments, and hence experimentally demonstrates the formation of a ferromagnetic phase for the DyAg$_2$ surface alloy at low sample temperature.

Based on the energy sequence of these spin split states, we assign the spectroscopic feature with positive spin polarization to the localized Dy 4f-derived states with majority spin character and the feature with negative spin polarization to the Dy 4f-derived states with minority spin-character. This is possible since the spin moments of electrons with majority character are aligned along the magnetization direction of the material thereby gaining energy, while the spin moments of electrons with minority character are oriented antiparallel to the magnetization and hence loose energy, i.e., the states of majority character are located at larger binding energies than those of minority character. Accordingly, all spectroscopic features with positive spin polarization are attributed to majority states and those with negative spin polarization to minority states.

Most importantly, these two states at binding energies smaller than $2\,$eV are the only non-dispersing states in the entire valence band structure of the DyAg$_2$ surface alloy up to a binding energy of E$_\mathrm{B}\approx 20\,$eV (see supplementary information). This is highly surprising since the Dy 4f states are usually observed in photoemission at significantly larger binding energies of at least $4\,$eV \cite{Lang1981,Gerken1983, Gerken1982}. This large binding energy is the result of strong electronic correlation effects of the localized 4f electrons leading to a reduced screening of the photohole in the final state of the photoemission process. In contrast, the binding energy of the Dy 4f states of the DyAg$_2$ surface alloy resembles more closely the one of 4f electrons of multi-compound materials consisting of rare earth atoms and light elements. For these systems, the electronic correlations are significantly reduced by a hybridization between the 4f electrons and the valence electrons of the light elements \cite{Neupane2013, Neupane2015, Danzenbcher2009}. Our findings hence point to a strong reduction of the electronic correlations of the localized Dy 4f electrons in the surface alloy.

Of similar importance is the spin contrast of the Dy-Ag hybrid surface state, which can be extracted from the exemplary CE maps in Fig.~3(a). Below E$_{\mathrm{F}}$, we find a ring-like emission feature (dashed yellow ring as guide to the eye in CE maps 3 and 4) with marginal majority spin character (red) which is surrounded by a hexagonal emission pattern (dashed yellow hexagon for guide to the eye in CE maps 2 and 3) with minority (blue) spin contrast. In the energy vs. momentum cut in the right half of Fig.~3(b), the minority spin polarization coincides reasonably well with the dispersion of the backfolded sp-bands of Ag indicating a spin selectivity of the surface umklapp process at the Dy-Ag layer. In addition, the Dy-Ag hybrid state itself exhibits only a negligible spin contrast in this binding energy range. Crucially, however, we find a ring-like emission with clear majority spin character in the high-energy wing of the Fermi distribution, i.e., at $E_\mathrm{B}\approx-0.1\;$eV, which becomes only visible when enhancing the intensity contrast of this CE map in Fig.~3(a), CE map 5. The energy vs. momentum cut in Fig.~3(b) also reveals a similar state with majority spin-polarization (red contrast) at the $\bar{\Gamma}^{\left(1\right)}$-point when normalizing the spin-resolved photoemission data by the Fermi-Dirac distribution (left side of Fig.~3b).  Hence, we propose that this band with majority spin character at the $\bar{\Gamma}^{\left(1\right)}$-point can be assigned to one spin branch of the hole-like Dy-Ag hybrid surface state.

To determine the origin of the spin polarized band structure of the Dy-Ag surface alloy, we performed density functional theory (DFT) calculations using the FLEUR ab initio package \cite{fleur}. For the calculations, we selected the LDA-VWN exchange and correlation functional \cite{Vosko.1980} and considered Hubbard parameter between $U=0\,$eV and $U=7\,$eV and an exchange interaction parameter $J=0.7\,$eV \cite{Anisimov.1997, Shick.1999}. Band structure calculations and the partial density of states of the Dy 4f electrons are shown in the supplementary information for different Hubbard parameters $U$. The Dy-Ag surface alloy was modeled by a supercell composed of $1$ Dy atom and $8$ Ag atoms resulting in one layer of the DyAg$_2$ surface alloy and $2$ layers of the Ag(111) substrate. The vertical relaxation of the Dy atom with respect to the Ag surface plane was fixed to $0.6\,$\AA. 

The calculated spin-resolved band structure along the $\bar{\Gamma}^{\left(1\right)}$~$\bar{\mathrm{M}}$ direction is shown in Fig.~3(c). We find an overall excellent agreement between the experimental band structure and the DFT+U calculations for $U=1.5\,$eV. Our band structure calculation predicts the existence of two sets of non-dispersing states of opposite spin character in the Dy-Ag valence band (labelled Dy 4f$^{\uparrow / \downarrow}$) with dominant 4f orbital character with can be attributed to the spin-split Dy 4f states. In addition, we find a Dy-Ag hybrid state at the $\bar{\Gamma}^{\left(1\right)}$-point of the SBZ (HSS$^{\uparrow / \downarrow}$) consisting of a mixture of Dy d- and the Ag p-orbitals (more details in the supporting information). Even more importantly, this hybrid surface state reveals a significant exchange splitting of $100\,$meV in the vicinity of the Fermi energy with an identical energy sequence of its spin-dependent branches as the localized Dy 4f states. The exchange splitting of this hybrid state is marked by a green arrow in Fig.~3c. This observation clearly points to an induced polarization of the Dy-Ag hybrid surface state electrons by the local magnetic moments of the Dy 4f electrons and hence to an RKKY-like exchange coupling in this material. The magnitude of the local magnetic moments of the Dy 4f states can also be extracted from our DFT+U simulation. We find a magnetic spin moment of $\mu_{\mathrm{Dy}}=4.3\,\mu_{\mathrm{B}}$ per Dy atom and a marginal magnetic moment on the Ag sites of the order of $\mu_{\mathrm{Ag}}=0.003\,\mu_{\mathrm{B}}$. 

Finally, we would like to point out that the best agreement between experiment and theory was obtained for a Hubbard parameter U$=1.5\,$eV. This value is surprisingly small compared to previous DFT+U calculations of bulk materials containing Dy \cite{Rahmanizadeh.2012,Raekers.2009} and hence points to a highly efficient screening of the Dy 4f electrons by the s-electrons of the host material Ag. The reduced electronic correlations are responsible for the exceptionally small binding energy of the Dy 4f states. This underlines the important role of the Dy-Ag interaction for the electronic correlations as well as for the spin-dependent and magnetic properties of the surface alloy.

In conclusion, our experimental and theoretical investigation of the spin-dependent electronic structure of a Dy-Ag surface alloy reveals clear spectroscopic indications for the formation of a low-dimensional ferromagnetic phase at low sample temperature of $40\,$K. We demonstrate a significant spin-splitting of the localized Dy 4f states as well as an exchange split hole-like Dy-Ag hybrid surface state in the center of the SBZ. This characteristic band structure shows that the long-range order of the localized Dy moment is mediated by the spin-polarized electrons of the Dy-Ag hybrid surface state. As a consequence, the magnitude of exchange splitting of the hybrid surface state crucially depends on the strength of the indirect, RKKY-like exchange coupling in the alloy layer as well as on the localized magnetic moments of the rare earth atoms. Considering the high tunability of binary surface alloys, our findings lay the foundation for tailoring and controlling the spin order and the spin-dependent charge carrier properties in low dimensional structures by surface alloying.

\begin{acknowledgments}
This work was funded by the Deutsche Forschungsgemeinschaft (DFG, German Research Foundation) - TRR 173 - 268565370 (projects A02, A03 and A09). The authors want to thank HZB for the allocation of synchrotron radiation beamtime. G.S., H.J. E. and B.S. thankfully acknowledge financial support from the Graduate School of Excellence MAINZ (Excellence Initiative DFG/GSC 266).
\end{acknowledgments}

%\bibliography{Literature}

\begin{thebibliography}{39}
\expandafter\ifx\csname natexlab\endcsname\relax\def\natexlab#1{#1}\fi
\expandafter\ifx\csname bibnamefont\endcsname\relax
  \def\bibnamefont#1{#1}\fi
\expandafter\ifx\csname bibfnamefont\endcsname\relax
  \def\bibfnamefont#1{#1}\fi
\expandafter\ifx\csname citenamefont\endcsname\relax
  \def\citenamefont#1{#1}\fi
\expandafter\ifx\csname url\endcsname\relax
  \def\url#1{\texttt{#1}}\fi
\expandafter\ifx\csname urlprefix\endcsname\relax\def\urlprefix{URL }\fi
\providecommand{\bibinfo}[2]{#2}
\providecommand{\eprint}[2][]{\url{#2}}

\bibitem[{\citenamefont{Wang et~al.}(2013)\citenamefont{Wang, Meric, Huang,
  Gao, Gao, Tran, Taniguchi, Watanabe, Campos, Muller et~al.}}]{Wang.2013}
\bibinfo{author}{\bibfnamefont{L.}~\bibnamefont{Wang}},
  \bibinfo{author}{\bibfnamefont{I.}~\bibnamefont{Meric}},
  \bibinfo{author}{\bibfnamefont{P.~Y.} \bibnamefont{Huang}},
  \bibinfo{author}{\bibfnamefont{Q.}~\bibnamefont{Gao}},
  \bibinfo{author}{\bibfnamefont{Y.}~\bibnamefont{Gao}},
  \bibinfo{author}{\bibfnamefont{H.}~\bibnamefont{Tran}},
  \bibinfo{author}{\bibfnamefont{T.}~\bibnamefont{Taniguchi}},
  \bibinfo{author}{\bibfnamefont{K.}~\bibnamefont{Watanabe}},
  \bibinfo{author}{\bibfnamefont{L.~M.} \bibnamefont{Campos}},
  \bibinfo{author}{\bibfnamefont{D.~A.} \bibnamefont{Muller}},
  \bibnamefont{et~al.}, \bibinfo{journal}{Science}
  \textbf{\bibinfo{volume}{342}}, \bibinfo{pages}{614} (\bibinfo{year}{2013}).

\bibitem[{\citenamefont{Dlubak et~al.}(2012)\citenamefont{Dlubak, Martin,
  Deranlot, Servet, Xavier, Mattana, Sprinkle, Berger, de~Heer, Petroff
  et~al.}}]{Dlubak.2012}
\bibinfo{author}{\bibfnamefont{B.}~\bibnamefont{Dlubak}},
  \bibinfo{author}{\bibfnamefont{M.-B.} \bibnamefont{Martin}},
  \bibinfo{author}{\bibfnamefont{C.}~\bibnamefont{Deranlot}},
  \bibinfo{author}{\bibfnamefont{B.}~\bibnamefont{Servet}},
  \bibinfo{author}{\bibfnamefont{S.}~\bibnamefont{Xavier}},
  \bibinfo{author}{\bibfnamefont{R.}~\bibnamefont{Mattana}},
  \bibinfo{author}{\bibfnamefont{M.}~\bibnamefont{Sprinkle}},
  \bibinfo{author}{\bibfnamefont{C.}~\bibnamefont{Berger}},
  \bibinfo{author}{\bibfnamefont{W.~A.} \bibnamefont{de~Heer}},
  \bibinfo{author}{\bibfnamefont{F.}~\bibnamefont{Petroff}},
  \bibnamefont{et~al.}, \bibinfo{journal}{Nat. Phys.}
  \textbf{\bibinfo{volume}{8}}, \bibinfo{pages}{557} (\bibinfo{year}{2012}).

\bibitem[{\citenamefont{Dankert and Dash}(2017)}]{Dankert.2017}
\bibinfo{author}{\bibfnamefont{A.}~\bibnamefont{Dankert}} \bibnamefont{and}
  \bibinfo{author}{\bibfnamefont{S.~P.} \bibnamefont{Dash}},
  \bibinfo{journal}{Nat. Commun.} \textbf{\bibinfo{volume}{8}},
  \bibinfo{pages}{16093} (\bibinfo{year}{2017}).

\bibitem[{\citenamefont{S{\'a}nchez et~al.}(2013)\citenamefont{S{\'a}nchez,
  Vila, Desfonds, Gambarelli, Attan{\'e}, de~Teresa, Mag{\'e}n, and
  Fert}}]{Sanchez.2013}
\bibinfo{author}{\bibfnamefont{J.~C.~R.} \bibnamefont{S{\'a}nchez}},
  \bibinfo{author}{\bibfnamefont{L.}~\bibnamefont{Vila}},
  \bibinfo{author}{\bibfnamefont{G.}~\bibnamefont{Desfonds}},
  \bibinfo{author}{\bibfnamefont{S.}~\bibnamefont{Gambarelli}},
  \bibinfo{author}{\bibfnamefont{J.~P.} \bibnamefont{Attan{\'e}}},
  \bibinfo{author}{\bibfnamefont{J.~M.} \bibnamefont{de~Teresa}},
  \bibinfo{author}{\bibfnamefont{C.}~\bibnamefont{Mag{\'e}n}},
  \bibnamefont{and} \bibinfo{author}{\bibfnamefont{A.}~\bibnamefont{Fert}},
  \bibinfo{journal}{{Nat. Commun.}} \textbf{\bibinfo{volume}{4}},
  \bibinfo{pages}{2944} (\bibinfo{year}{2013}).

\bibitem[{\citenamefont{Manchon et~al.}(2015)\citenamefont{Manchon, Koo, Nitta,
  Frolov, and Duine}}]{Manchon.2015}
\bibinfo{author}{\bibfnamefont{A.}~\bibnamefont{Manchon}},
  \bibinfo{author}{\bibfnamefont{H.~C.} \bibnamefont{Koo}},
  \bibinfo{author}{\bibfnamefont{J.}~\bibnamefont{Nitta}},
  \bibinfo{author}{\bibfnamefont{S.~M.} \bibnamefont{Frolov}},
  \bibnamefont{and} \bibinfo{author}{\bibfnamefont{R.~A.} \bibnamefont{Duine}},
  \bibinfo{journal}{{Nat. Mater.}} \textbf{\bibinfo{volume}{14}},
  \bibinfo{pages}{871} (\bibinfo{year}{2015}).

\bibitem[{\citenamefont{Ast et~al.}(2007)\citenamefont{Ast, Henk, Ernst,
  Moreschini, Falub, Pacil{\'e}, Bruno, Kern, and Grioni}}]{Ast.2007}
\bibinfo{author}{\bibfnamefont{C.~R.} \bibnamefont{Ast}},
  \bibinfo{author}{\bibfnamefont{J.}~\bibnamefont{Henk}},
  \bibinfo{author}{\bibfnamefont{A.}~\bibnamefont{Ernst}},
  \bibinfo{author}{\bibfnamefont{L.}~\bibnamefont{Moreschini}},
  \bibinfo{author}{\bibfnamefont{M.~C.} \bibnamefont{Falub}},
  \bibinfo{author}{\bibfnamefont{D.}~\bibnamefont{Pacil{\'e}}},
  \bibinfo{author}{\bibfnamefont{P.}~\bibnamefont{Bruno}},
  \bibinfo{author}{\bibfnamefont{K.}~\bibnamefont{Kern}}, \bibnamefont{and}
  \bibinfo{author}{\bibfnamefont{M.}~\bibnamefont{Grioni}},
  \bibinfo{journal}{Phys. Rev. Lett.} \textbf{\bibinfo{volume}{98}},
  \bibinfo{pages}{186807} (\bibinfo{year}{2007}).

\bibitem[{\citenamefont{Bentmann et~al.}(2009)\citenamefont{Bentmann, Forster,
  Bihlmayer, Chulkov, Moreschini, Grioni, and Reinert}}]{Bentmann.2009}
\bibinfo{author}{\bibfnamefont{H.}~\bibnamefont{Bentmann}},
  \bibinfo{author}{\bibfnamefont{F.}~\bibnamefont{Forster}},
  \bibinfo{author}{\bibfnamefont{G.}~\bibnamefont{Bihlmayer}},
  \bibinfo{author}{\bibfnamefont{E.~V.} \bibnamefont{Chulkov}},
  \bibinfo{author}{\bibfnamefont{L.}~\bibnamefont{Moreschini}},
  \bibinfo{author}{\bibfnamefont{M.}~\bibnamefont{Grioni}}, \bibnamefont{and}
  \bibinfo{author}{\bibfnamefont{F.}~\bibnamefont{Reinert}},
  \bibinfo{journal}{Europhys. Lett.} \textbf{\bibinfo{volume}{87}},
  \bibinfo{pages}{37003} (\bibinfo{year}{2009}).

\bibitem[{\citenamefont{Moreschini et~al.}(2009)\citenamefont{Moreschini,
  Bendounan, Bentmann, Assig, Kern, Reinert, Henk, Ast, and
  Grioni}}]{Moreschini.2009}
\bibinfo{author}{\bibfnamefont{L.}~\bibnamefont{Moreschini}},
  \bibinfo{author}{\bibfnamefont{A.}~\bibnamefont{Bendounan}},
  \bibinfo{author}{\bibfnamefont{H.}~\bibnamefont{Bentmann}},
  \bibinfo{author}{\bibfnamefont{M.}~\bibnamefont{Assig}},
  \bibinfo{author}{\bibfnamefont{K.}~\bibnamefont{Kern}},
  \bibinfo{author}{\bibfnamefont{F.}~\bibnamefont{Reinert}},
  \bibinfo{author}{\bibfnamefont{J.}~\bibnamefont{Henk}},
  \bibinfo{author}{\bibfnamefont{C.~R.} \bibnamefont{Ast}}, \bibnamefont{and}
  \bibinfo{author}{\bibfnamefont{M.}~\bibnamefont{Grioni}},
  \bibinfo{journal}{Phys. Rev. B} \textbf{\bibinfo{volume}{80}}
  (\bibinfo{year}{2009}).

\bibitem[{\citenamefont{Meier et~al.}(2011)\citenamefont{Meier, Petrov,
  Mirhosseini, Patthey, Henk, Osterwalder, and Dil}}]{Meier.2011}
\bibinfo{author}{\bibfnamefont{F.}~\bibnamefont{Meier}},
  \bibinfo{author}{\bibfnamefont{V.}~\bibnamefont{Petrov}},
  \bibinfo{author}{\bibfnamefont{H.}~\bibnamefont{Mirhosseini}},
  \bibinfo{author}{\bibfnamefont{L.}~\bibnamefont{Patthey}},
  \bibinfo{author}{\bibfnamefont{J.}~\bibnamefont{Henk}},
  \bibinfo{author}{\bibfnamefont{J.}~\bibnamefont{Osterwalder}},
  \bibnamefont{and} \bibinfo{author}{\bibfnamefont{J.~H.} \bibnamefont{Dil}},
  \bibinfo{journal}{J. Phys.: Condens. Matter} \textbf{\bibinfo{volume}{23}},
  \bibinfo{pages}{072207} (\bibinfo{year}{2011}).

\bibitem[{\citenamefont{Bihlmayer et~al.}(2007)\citenamefont{Bihlmayer,
  Bl{\"u}gel, and Chulkov}}]{Bihlmayer.2007}
\bibinfo{author}{\bibfnamefont{G.}~\bibnamefont{Bihlmayer}},
  \bibinfo{author}{\bibfnamefont{S.}~\bibnamefont{Bl{\"u}gel}},
  \bibnamefont{and} \bibinfo{author}{\bibfnamefont{E.~V.}
  \bibnamefont{Chulkov}}, \bibinfo{journal}{Phys. Rev. B}
  \textbf{\bibinfo{volume}{75}}, \bibinfo{pages}{195414}
  (\bibinfo{year}{2007}).

\bibitem[{\citenamefont{Maniraj et~al.}(2018)\citenamefont{Maniraj, Jungkenn,
  Shi, Emmerich, Lyu, Kollamana, Wei, Yan, Cinchetti, Mathias
  et~al.}}]{Maniraj.2018}
\bibinfo{author}{\bibfnamefont{M.}~\bibnamefont{Maniraj}},
  \bibinfo{author}{\bibfnamefont{D.}~\bibnamefont{Jungkenn}},
  \bibinfo{author}{\bibfnamefont{W.}~\bibnamefont{Shi}},
  \bibinfo{author}{\bibfnamefont{S.}~\bibnamefont{Emmerich}},
  \bibinfo{author}{\bibfnamefont{L.}~\bibnamefont{Lyu}},
  \bibinfo{author}{\bibfnamefont{J.}~\bibnamefont{Kollamana}},
  \bibinfo{author}{\bibfnamefont{Z.}~\bibnamefont{Wei}},
  \bibinfo{author}{\bibfnamefont{B.}~\bibnamefont{Yan}},
  \bibinfo{author}{\bibfnamefont{M.}~\bibnamefont{Cinchetti}},
  \bibinfo{author}{\bibfnamefont{S.}~\bibnamefont{Mathias}},
  \bibnamefont{et~al.}, \bibinfo{journal}{Phys. Rev. B}
  \textbf{\bibinfo{volume}{98}}, \bibinfo{pages}{205419}
  (\bibinfo{year}{2018}).

\bibitem[{\citenamefont{Petersen and Hedeg{\aa}rd}(2000)}]{Petersen.2000}
\bibinfo{author}{\bibfnamefont{L.}~\bibnamefont{Petersen}} \bibnamefont{and}
  \bibinfo{author}{\bibfnamefont{P.}~\bibnamefont{Hedeg{\aa}rd}},
  \bibinfo{journal}{Surf. Sci.} \textbf{\bibinfo{volume}{459}},
  \bibinfo{pages}{49} (\bibinfo{year}{2000}).

\bibitem[{\citenamefont{Gierz et~al.}(2011)\citenamefont{Gierz, Meier, Dil,
  Kern, and Ast}}]{Gierz.2011}
\bibinfo{author}{\bibfnamefont{I.}~\bibnamefont{Gierz}},
  \bibinfo{author}{\bibfnamefont{F.}~\bibnamefont{Meier}},
  \bibinfo{author}{\bibfnamefont{J.~H.} \bibnamefont{Dil}},
  \bibinfo{author}{\bibfnamefont{K.}~\bibnamefont{Kern}}, \bibnamefont{and}
  \bibinfo{author}{\bibfnamefont{C.~R.} \bibnamefont{Ast}},
  \bibinfo{journal}{Phys. Rev. B} \textbf{\bibinfo{volume}{83}},
  \bibinfo{pages}{195122} (\bibinfo{year}{2011}).

\bibitem[{\citenamefont{Gierz et~al.}(2010)\citenamefont{Gierz,
  Stadtm{\"u}ller, Vuorinen, Lindroos, Meier, Dil, Kern, and Ast}}]{Gierz.2010}
\bibinfo{author}{\bibfnamefont{I.}~\bibnamefont{Gierz}},
  \bibinfo{author}{\bibfnamefont{B.}~\bibnamefont{Stadtm{\"u}ller}},
  \bibinfo{author}{\bibfnamefont{J.}~\bibnamefont{Vuorinen}},
  \bibinfo{author}{\bibfnamefont{M.}~\bibnamefont{Lindroos}},
  \bibinfo{author}{\bibfnamefont{F.}~\bibnamefont{Meier}},
  \bibinfo{author}{\bibfnamefont{J.~H.} \bibnamefont{Dil}},
  \bibinfo{author}{\bibfnamefont{K.}~\bibnamefont{Kern}}, \bibnamefont{and}
  \bibinfo{author}{\bibfnamefont{C.~R.} \bibnamefont{Ast}},
  \bibinfo{journal}{Phys. Rev. B} \textbf{\bibinfo{volume}{81}}
  (\bibinfo{year}{2010}).

\bibitem[{\citenamefont{Stadtm{\"u}ller
  et~al.}(2016)\citenamefont{Stadtm{\"u}ller, Seidel, Haag, Grad, Tusche, {van
  Straaten}, Franke, Kirschner, Kumpf, Cinchetti et~al.}}]{Stadtmuller.2016}
\bibinfo{author}{\bibfnamefont{B.}~\bibnamefont{Stadtm{\"u}ller}},
  \bibinfo{author}{\bibfnamefont{J.}~\bibnamefont{Seidel}},
  \bibinfo{author}{\bibfnamefont{N.}~\bibnamefont{Haag}},
  \bibinfo{author}{\bibfnamefont{L.}~\bibnamefont{Grad}},
  \bibinfo{author}{\bibfnamefont{C.}~\bibnamefont{Tusche}},
  \bibinfo{author}{\bibfnamefont{G.}~\bibnamefont{{van Straaten}}},
  \bibinfo{author}{\bibfnamefont{M.}~\bibnamefont{Franke}},
  \bibinfo{author}{\bibfnamefont{J.}~\bibnamefont{Kirschner}},
  \bibinfo{author}{\bibfnamefont{C.}~\bibnamefont{Kumpf}},
  \bibinfo{author}{\bibfnamefont{M.}~\bibnamefont{Cinchetti}},
  \bibnamefont{et~al.}, \bibinfo{journal}{Phys. Rev. Lett.}
  \textbf{\bibinfo{volume}{117}}, \bibinfo{pages}{096805}
  (\bibinfo{year}{2016}).

\bibitem[{\citenamefont{Friedrich et~al.}(2017)\citenamefont{Friedrich, Caciuc,
  Bihlmayer, Atodiresei, and Bl{\"u}gel}}]{Friedrich.2017}
\bibinfo{author}{\bibfnamefont{R.}~\bibnamefont{Friedrich}},
  \bibinfo{author}{\bibfnamefont{V.}~\bibnamefont{Caciuc}},
  \bibinfo{author}{\bibfnamefont{G.}~\bibnamefont{Bihlmayer}},
  \bibinfo{author}{\bibfnamefont{N.}~\bibnamefont{Atodiresei}},
  \bibnamefont{and}
  \bibinfo{author}{\bibfnamefont{S.}~\bibnamefont{Bl{\"u}gel}},
  \bibinfo{journal}{New J. Phys.} \textbf{\bibinfo{volume}{19}},
  \bibinfo{pages}{043017} (\bibinfo{year}{2017}).

\bibitem[{\citenamefont{Ormaza et~al.}(2016)\citenamefont{Ormaza, Fernandez,
  Ilyn, Magana, Xu, Verstraete, Gastaldo, Valbuena, Gargiani, Mugarza
  et~al.}}]{Ormaza.2016}
\bibinfo{author}{\bibfnamefont{M.}~\bibnamefont{Ormaza}},
  \bibinfo{author}{\bibfnamefont{L.}~\bibnamefont{Fernandez}},
  \bibinfo{author}{\bibfnamefont{M.}~\bibnamefont{Ilyn}},
  \bibinfo{author}{\bibfnamefont{A.}~\bibnamefont{Magana}},
  \bibinfo{author}{\bibfnamefont{B.}~\bibnamefont{Xu}},
  \bibinfo{author}{\bibfnamefont{M.~J.} \bibnamefont{Verstraete}},
  \bibinfo{author}{\bibfnamefont{M.}~\bibnamefont{Gastaldo}},
  \bibinfo{author}{\bibfnamefont{M.~A.} \bibnamefont{Valbuena}},
  \bibinfo{author}{\bibfnamefont{P.}~\bibnamefont{Gargiani}},
  \bibinfo{author}{\bibfnamefont{A.}~\bibnamefont{Mugarza}},
  \bibnamefont{et~al.}, \bibinfo{journal}{Nano Lett.} p. \bibinfo{pages}{4230}
  (\bibinfo{year}{2016}).

\bibitem[{\citenamefont{Correa et~al.}(2016)\citenamefont{Correa, Xu,
  Verstraete, and Vitali}}]{Correa.2016}
\bibinfo{author}{\bibfnamefont{A.}~\bibnamefont{Correa}},
  \bibinfo{author}{\bibfnamefont{B.}~\bibnamefont{Xu}},
  \bibinfo{author}{\bibfnamefont{M.~J.} \bibnamefont{Verstraete}},
  \bibnamefont{and} \bibinfo{author}{\bibfnamefont{L.}~\bibnamefont{Vitali}},
  \bibinfo{journal}{Nanoscale} \textbf{\bibinfo{volume}{8}},
  \bibinfo{pages}{19148} (\bibinfo{year}{2016}).

\bibitem[{\citenamefont{Medjanik et~al.}(2017)\citenamefont{Medjanik,
  Fedchenko, Chernov, Kutnyakhov, Ellguth, Oelsner, Schonhense, Peixoto, Lutz,
  Min et~al.}}]{Medjanik.2017}
\bibinfo{author}{\bibfnamefont{K.}~\bibnamefont{Medjanik}},
  \bibinfo{author}{\bibfnamefont{O.}~\bibnamefont{Fedchenko}},
  \bibinfo{author}{\bibfnamefont{S.}~\bibnamefont{Chernov}},
  \bibinfo{author}{\bibfnamefont{D.}~\bibnamefont{Kutnyakhov}},
  \bibinfo{author}{\bibfnamefont{M.}~\bibnamefont{Ellguth}},
  \bibinfo{author}{\bibfnamefont{A.}~\bibnamefont{Oelsner}},
  \bibinfo{author}{\bibfnamefont{B.}~\bibnamefont{Schonhense}},
  \bibinfo{author}{\bibfnamefont{T.~R.~F.} \bibnamefont{Peixoto}},
  \bibinfo{author}{\bibfnamefont{P.}~\bibnamefont{Lutz}},
  \bibinfo{author}{\bibfnamefont{C.-H.} \bibnamefont{Min}},
  \bibnamefont{et~al.}, \bibinfo{journal}{Nat. Mater.} pp.
  \bibinfo{pages}{pages 615--621} (\bibinfo{year}{2017}).

\bibitem[{\citenamefont{Sch{\"o}nhense
  et~al.}(2017)\citenamefont{Sch{\"o}nhense, Medjanik, Chernov, Kutnyakhov,
  Fedchenko, Ellguth, Vasilyev, Zaporozhchenko-Zymakov{\'a}, Panzer, Oelsner
  et~al.}}]{Schonhense.2017}
\bibinfo{author}{\bibfnamefont{G.}~\bibnamefont{Sch{\"o}nhense}},
  \bibinfo{author}{\bibfnamefont{K.}~\bibnamefont{Medjanik}},
  \bibinfo{author}{\bibfnamefont{S.}~\bibnamefont{Chernov}},
  \bibinfo{author}{\bibfnamefont{D.}~\bibnamefont{Kutnyakhov}},
  \bibinfo{author}{\bibfnamefont{O.}~\bibnamefont{Fedchenko}},
  \bibinfo{author}{\bibfnamefont{M.}~\bibnamefont{Ellguth}},
  \bibinfo{author}{\bibfnamefont{D.}~\bibnamefont{Vasilyev}},
  \bibinfo{author}{\bibfnamefont{A.}~\bibnamefont{Zaporozhchenko-Zymakov{\'a}}},
  \bibinfo{author}{\bibfnamefont{D.}~\bibnamefont{Panzer}},
  \bibinfo{author}{\bibfnamefont{A.}~\bibnamefont{Oelsner}},
  \bibnamefont{et~al.}, \bibinfo{journal}{Ultramicroscopy}
  \textbf{\bibinfo{volume}{183}}, \bibinfo{pages}{19} (\bibinfo{year}{2017}).

\bibitem[{\citenamefont{Kolbe et~al.}(2011)\citenamefont{Kolbe, Lushchyk,
  Petereit, Elmers, Sch{\"o}nhense, Oelsner, Tusche, and
  Kirschner}}]{Kolbe.2011}
\bibinfo{author}{\bibfnamefont{M.}~\bibnamefont{Kolbe}},
  \bibinfo{author}{\bibfnamefont{P.}~\bibnamefont{Lushchyk}},
  \bibinfo{author}{\bibfnamefont{B.}~\bibnamefont{Petereit}},
  \bibinfo{author}{\bibfnamefont{H.~J.} \bibnamefont{Elmers}},
  \bibinfo{author}{\bibfnamefont{G.}~\bibnamefont{Sch{\"o}nhense}},
  \bibinfo{author}{\bibfnamefont{A.}~\bibnamefont{Oelsner}},
  \bibinfo{author}{\bibfnamefont{C.}~\bibnamefont{Tusche}}, \bibnamefont{and}
  \bibinfo{author}{\bibfnamefont{J.}~\bibnamefont{Kirschner}},
  \bibinfo{journal}{Phys. Rev. Lett.} \textbf{\bibinfo{volume}{107}},
  \bibinfo{pages}{207601} (\bibinfo{year}{2011}), ISSN
  \bibinfo{issn}{0031-9007}.

\bibitem[{\citenamefont{Kutnyakhov et~al.}(2013)\citenamefont{Kutnyakhov,
  Lushchyk, Fognini, Perriard, Kolbe, Medjanik, Fedchenko, Nepijko, Elmers,
  Salvatella et~al.}}]{Kutnyakhov.2013}
\bibinfo{author}{\bibfnamefont{D.}~\bibnamefont{Kutnyakhov}},
  \bibinfo{author}{\bibfnamefont{P.}~\bibnamefont{Lushchyk}},
  \bibinfo{author}{\bibfnamefont{A.}~\bibnamefont{Fognini}},
  \bibinfo{author}{\bibfnamefont{D.}~\bibnamefont{Perriard}},
  \bibinfo{author}{\bibfnamefont{M.}~\bibnamefont{Kolbe}},
  \bibinfo{author}{\bibfnamefont{K.}~\bibnamefont{Medjanik}},
  \bibinfo{author}{\bibfnamefont{E.}~\bibnamefont{Fedchenko}},
  \bibinfo{author}{\bibfnamefont{S.~A.} \bibnamefont{Nepijko}},
  \bibinfo{author}{\bibfnamefont{H.~J.} \bibnamefont{Elmers}},
  \bibinfo{author}{\bibfnamefont{G.}~\bibnamefont{Salvatella}},
  \bibnamefont{et~al.}, \bibinfo{journal}{Ultramicroscopy}
  \textbf{\bibinfo{volume}{130}}, \bibinfo{pages}{63} (\bibinfo{year}{2013}).

\bibitem[{\citenamefont{{Simon Moser}}(2017)}]{SimonMoser.2017}
\bibinfo{author}{\bibnamefont{{Simon Moser}}}, \bibinfo{journal}{J. Electron
  Spectrosc. Relat. Phenom.} \textbf{\bibinfo{volume}{214}},
  \bibinfo{pages}{29} (\bibinfo{year}{2017}).

\bibitem[{\citenamefont{Anderson and Lapeyre}(1976)}]{Anderson.1976}
\bibinfo{author}{\bibfnamefont{J.}~\bibnamefont{Anderson}} \bibnamefont{and}
  \bibinfo{author}{\bibfnamefont{G.~J.} \bibnamefont{Lapeyre}},
  \bibinfo{journal}{Phys. Rev. Lett.} \textbf{\bibinfo{volume}{36}},
  \bibinfo{pages}{376} (\bibinfo{year}{1976}).

\bibitem[{\citenamefont{{\"U}nal et~al.}(2012)\citenamefont{{\"U}nal,
  Winkelmann, Tusche, Bisio, Ellguth, Chiang, Henk, and Kirschner}}]{Unal.2012}
\bibinfo{author}{\bibfnamefont{A.~A.} \bibnamefont{{\"U}nal}},
  \bibinfo{author}{\bibfnamefont{A.}~\bibnamefont{Winkelmann}},
  \bibinfo{author}{\bibfnamefont{C.}~\bibnamefont{Tusche}},
  \bibinfo{author}{\bibfnamefont{F.}~\bibnamefont{Bisio}},
  \bibinfo{author}{\bibfnamefont{M.}~\bibnamefont{Ellguth}},
  \bibinfo{author}{\bibfnamefont{C.-T.} \bibnamefont{Chiang}},
  \bibinfo{author}{\bibfnamefont{J.}~\bibnamefont{Henk}}, \bibnamefont{and}
  \bibinfo{author}{\bibfnamefont{J.}~\bibnamefont{Kirschner}},
  \bibinfo{journal}{Phys. Rev. B} \textbf{\bibinfo{volume}{86}},
  \bibinfo{pages}{125447} (\bibinfo{year}{2012}).

\bibitem[{\citenamefont{Giovanelli et~al.}(2013)\citenamefont{Giovanelli,
  Bocquet, Amsalem, Lee, Abel, Clair, Koudia, Faury, Petaccia, Topwal
  et~al.}}]{Giovanelli.2013}
\bibinfo{author}{\bibfnamefont{L.}~\bibnamefont{Giovanelli}},
  \bibinfo{author}{\bibfnamefont{F.~C.} \bibnamefont{Bocquet}},
  \bibinfo{author}{\bibfnamefont{P.}~\bibnamefont{Amsalem}},
  \bibinfo{author}{\bibfnamefont{H.-L.} \bibnamefont{Lee}},
  \bibinfo{author}{\bibfnamefont{M.}~\bibnamefont{Abel}},
  \bibinfo{author}{\bibfnamefont{S.}~\bibnamefont{Clair}},
  \bibinfo{author}{\bibfnamefont{M.}~\bibnamefont{Koudia}},
  \bibinfo{author}{\bibfnamefont{T.}~\bibnamefont{Faury}},
  \bibinfo{author}{\bibfnamefont{L.}~\bibnamefont{Petaccia}},
  \bibinfo{author}{\bibfnamefont{D.}~\bibnamefont{Topwal}},
  \bibnamefont{et~al.}, \bibinfo{journal}{Phys. Rev. B}
  \textbf{\bibinfo{volume}{87}}, \bibinfo{pages}{035413}
  (\bibinfo{year}{2013}).

\bibitem[{\citenamefont{St{\"o}ckl et~al.}(2018)\citenamefont{St{\"o}ckl,
  Jurenkow, Gro{\ss}mann, Cinchetti, Stadtm{\"u}ller, and
  Aeschlimann}}]{Stockl.2018}
\bibinfo{author}{\bibfnamefont{J.}~\bibnamefont{St{\"o}ckl}},
  \bibinfo{author}{\bibfnamefont{A.}~\bibnamefont{Jurenkow}},
  \bibinfo{author}{\bibfnamefont{N.}~\bibnamefont{Gro{\ss}mann}},
  \bibinfo{author}{\bibfnamefont{M.}~\bibnamefont{Cinchetti}},
  \bibinfo{author}{\bibfnamefont{B.}~\bibnamefont{Stadtm{\"u}ller}},
  \bibnamefont{and}
  \bibinfo{author}{\bibfnamefont{M.}~\bibnamefont{Aeschlimann}},
  \bibinfo{journal}{J. Phys. Chem. C} \textbf{\bibinfo{volume}{122}},
  \bibinfo{pages}{6585} (\bibinfo{year}{2018}).

\bibitem[{\citenamefont{Lang et~al.}(1981)\citenamefont{Lang, Baer, and
  Cox}}]{Lang1981}
\bibinfo{author}{\bibfnamefont{J.~K.} \bibnamefont{Lang}},
  \bibinfo{author}{\bibfnamefont{Y.}~\bibnamefont{Baer}}, \bibnamefont{and}
  \bibinfo{author}{\bibfnamefont{P.~A.} \bibnamefont{Cox}}, \bibinfo{journal}{J
  Phys Condens Matter} \textbf{\bibinfo{volume}{11}}, \bibinfo{pages}{121}
  (\bibinfo{year}{1981}).

\bibitem[{\citenamefont{Gerken}(1983)}]{Gerken1983}
\bibinfo{author}{\bibfnamefont{F.}~\bibnamefont{Gerken}}, \bibinfo{journal}{J
  Phys Condens Matter} \textbf{\bibinfo{volume}{13}}, \bibinfo{pages}{703}
  (\bibinfo{year}{1983}).

\bibitem[{\citenamefont{Gerken et~al.}(1982)\citenamefont{Gerken, Barth,
  Kammerer, Johansson, and Flodstr\"{o}m}}]{Gerken1982}
\bibinfo{author}{\bibfnamefont{F.}~\bibnamefont{Gerken}},
  \bibinfo{author}{\bibfnamefont{J.}~\bibnamefont{Barth}},
  \bibinfo{author}{\bibfnamefont{R.}~\bibnamefont{Kammerer}},
  \bibinfo{author}{\bibfnamefont{L.}~\bibnamefont{Johansson}},
  \bibnamefont{and}
  \bibinfo{author}{\bibfnamefont{A.}~\bibnamefont{Flodstr\"{o}m}},
  \bibinfo{journal}{Surf Sci} \textbf{\bibinfo{volume}{117}},
  \bibinfo{pages}{468} (\bibinfo{year}{1982}).

\bibitem[{\citenamefont{Neupane et~al.}(2013)\citenamefont{Neupane, Alidoust,
  Xu, Kondo, Ishida, Kim, Liu, Belopolski, Jo, Chang et~al.}}]{Neupane2013}
\bibinfo{author}{\bibfnamefont{M.}~\bibnamefont{Neupane}},
  \bibinfo{author}{\bibfnamefont{N.}~\bibnamefont{Alidoust}},
  \bibinfo{author}{\bibfnamefont{S.-Y.} \bibnamefont{Xu}},
  \bibinfo{author}{\bibfnamefont{T.}~\bibnamefont{Kondo}},
  \bibinfo{author}{\bibfnamefont{Y.}~\bibnamefont{Ishida}},
  \bibinfo{author}{\bibfnamefont{D.~J.} \bibnamefont{Kim}},
  \bibinfo{author}{\bibfnamefont{C.}~\bibnamefont{Liu}},
  \bibinfo{author}{\bibfnamefont{I.}~\bibnamefont{Belopolski}},
  \bibinfo{author}{\bibfnamefont{Y.~J.} \bibnamefont{Jo}},
  \bibinfo{author}{\bibfnamefont{T.-R.} \bibnamefont{Chang}},
  \bibnamefont{et~al.}, \bibinfo{journal}{Nat Commun}
  \textbf{\bibinfo{volume}{4}} (\bibinfo{year}{2013}).

\bibitem[{\citenamefont{Neupane et~al.}(2015)\citenamefont{Neupane, Xu,
  Alidoust, Bian, Kim, Liu, Belopolski, Chang, Jeng, Durakiewicz
  et~al.}}]{Neupane2015}
\bibinfo{author}{\bibfnamefont{M.}~\bibnamefont{Neupane}},
  \bibinfo{author}{\bibfnamefont{S.-Y.} \bibnamefont{Xu}},
  \bibinfo{author}{\bibfnamefont{N.}~\bibnamefont{Alidoust}},
  \bibinfo{author}{\bibfnamefont{G.}~\bibnamefont{Bian}},
  \bibinfo{author}{\bibfnamefont{D.}~\bibnamefont{Kim}},
  \bibinfo{author}{\bibfnamefont{C.}~\bibnamefont{Liu}},
  \bibinfo{author}{\bibfnamefont{I.}~\bibnamefont{Belopolski}},
  \bibinfo{author}{\bibfnamefont{T.-R.} \bibnamefont{Chang}},
  \bibinfo{author}{\bibfnamefont{H.-T.} \bibnamefont{Jeng}},
  \bibinfo{author}{\bibfnamefont{T.}~\bibnamefont{Durakiewicz}},
  \bibnamefont{et~al.}, \bibinfo{journal}{Phys. Rev. Lett.}
  \textbf{\bibinfo{volume}{114}} (\bibinfo{year}{2015}).

\bibitem[{\citenamefont{Danzenb\"{a}cher
  et~al.}(2009)\citenamefont{Danzenb\"{a}cher, Vyalikh, Kucherenko, Kade,
  Laubschat, Caroca-Canales, Krellner, Geibel, Fedorov, Dessau
  et~al.}}]{Danzenbcher2009}
\bibinfo{author}{\bibfnamefont{S.}~\bibnamefont{Danzenb\"{a}cher}},
  \bibinfo{author}{\bibfnamefont{D.~V.} \bibnamefont{Vyalikh}},
  \bibinfo{author}{\bibfnamefont{Y.}~\bibnamefont{Kucherenko}},
  \bibinfo{author}{\bibfnamefont{A.}~\bibnamefont{Kade}},
  \bibinfo{author}{\bibfnamefont{C.}~\bibnamefont{Laubschat}},
  \bibinfo{author}{\bibfnamefont{N.}~\bibnamefont{Caroca-Canales}},
  \bibinfo{author}{\bibfnamefont{C.}~\bibnamefont{Krellner}},
  \bibinfo{author}{\bibfnamefont{C.}~\bibnamefont{Geibel}},
  \bibinfo{author}{\bibfnamefont{A.~V.} \bibnamefont{Fedorov}},
  \bibinfo{author}{\bibfnamefont{D.~S.} \bibnamefont{Dessau}},
  \bibnamefont{et~al.}, \bibinfo{journal}{Phys. Rev. Lett.}
  \textbf{\bibinfo{volume}{102}} (\bibinfo{year}{2009}).

\bibitem[{fle()}]{fleur}
\bibinfo{howpublished}{\url{http://www.flapw.de}}.

\bibitem[{\citenamefont{Vosko et~al.}(1980)\citenamefont{Vosko, Wilk, and
  Nusair}}]{Vosko.1980}
\bibinfo{author}{\bibfnamefont{S.~H.} \bibnamefont{Vosko}},
  \bibinfo{author}{\bibfnamefont{L.}~\bibnamefont{Wilk}}, \bibnamefont{and}
  \bibinfo{author}{\bibfnamefont{M.}~\bibnamefont{Nusair}},
  \bibinfo{journal}{Can. J. Phys.} \textbf{\bibinfo{volume}{58}},
  \bibinfo{pages}{1200} (\bibinfo{year}{1980}), ISSN \bibinfo{issn}{0008-4204}.

\bibitem[{\citenamefont{Anisimov et~al.}(1997)\citenamefont{Anisimov,
  Aryasetiawan, and Lichtenstein}}]{Anisimov.1997}
\bibinfo{author}{\bibfnamefont{V.~I.} \bibnamefont{Anisimov}},
  \bibinfo{author}{\bibfnamefont{F.}~\bibnamefont{Aryasetiawan}},
  \bibnamefont{and} \bibinfo{author}{\bibfnamefont{A.~I.}
  \bibnamefont{Lichtenstein}}, \bibinfo{journal}{J. Phys. Condens. Matter}
  \textbf{\bibinfo{volume}{9}}, \bibinfo{pages}{767} (\bibinfo{year}{1997}),
  ISSN \bibinfo{issn}{0953-8984}.

\bibitem[{\citenamefont{Shick et~al.}(1999)\citenamefont{Shick, Liechtenstein,
  and Pickett}}]{Shick.1999}
\bibinfo{author}{\bibfnamefont{A.~B.} \bibnamefont{Shick}},
  \bibinfo{author}{\bibfnamefont{A.~I.} \bibnamefont{Liechtenstein}},
  \bibnamefont{and} \bibinfo{author}{\bibfnamefont{W.~E.}
  \bibnamefont{Pickett}}, \bibinfo{journal}{Phys. Rev. B}
  \textbf{\bibinfo{volume}{60}}, \bibinfo{pages}{10763} (\bibinfo{year}{1999}),
  ISSN \bibinfo{issn}{1098-0121}.

\bibitem[{\citenamefont{Rahmanizadeh et~al.}(2012)\citenamefont{Rahmanizadeh,
  Bihlmayer, Luysberg, and Bl{\"u}gel}}]{Rahmanizadeh.2012}
\bibinfo{author}{\bibfnamefont{K.}~\bibnamefont{Rahmanizadeh}},
  \bibinfo{author}{\bibfnamefont{G.}~\bibnamefont{Bihlmayer}},
  \bibinfo{author}{\bibfnamefont{M.}~\bibnamefont{Luysberg}}, \bibnamefont{and}
  \bibinfo{author}{\bibfnamefont{S.}~\bibnamefont{Bl{\"u}gel}},
  \bibinfo{journal}{Phys. Rev. B} \textbf{\bibinfo{volume}{85}},
  \bibinfo{pages}{075314} (\bibinfo{year}{2012}).

\bibitem[{\citenamefont{Raekers et~al.}(2009)\citenamefont{Raekers, Kuepper,
  Bartkowski, Prinz, Postnikov, Potzger, Zhou, Arulraj, St{\"u}{\ss}er, Uecker
  et~al.}}]{Raekers.2009}
\bibinfo{author}{\bibfnamefont{M.}~\bibnamefont{Raekers}},
  \bibinfo{author}{\bibfnamefont{K.}~\bibnamefont{Kuepper}},
  \bibinfo{author}{\bibfnamefont{S.}~\bibnamefont{Bartkowski}},
  \bibinfo{author}{\bibfnamefont{M.}~\bibnamefont{Prinz}},
  \bibinfo{author}{\bibfnamefont{A.~V.} \bibnamefont{Postnikov}},
  \bibinfo{author}{\bibfnamefont{K.}~\bibnamefont{Potzger}},
  \bibinfo{author}{\bibfnamefont{S.}~\bibnamefont{Zhou}},
  \bibinfo{author}{\bibfnamefont{A.}~\bibnamefont{Arulraj}},
  \bibinfo{author}{\bibfnamefont{N.}~\bibnamefont{St{\"u}{\ss}er}},
  \bibinfo{author}{\bibfnamefont{R.}~\bibnamefont{Uecker}},
  \bibnamefont{et~al.}, \bibinfo{journal}{Phys. Rev. B}
  \textbf{\bibinfo{volume}{79}}, \bibinfo{pages}{125114}
  (\bibinfo{year}{2009}).

\end{thebibliography}
%\includepdf[pages={{},1,{}}]{supplementary_information.pdf}
\end{document}

\title{Exchange Splitting of a Hybrid Surface State and Ferromagnetic Order in a 2D Surface Alloy\\Supplemental Material}

\author{Johannes Knippertz}
\email[]{jseidel@rhrk.uni-kl.de}
\affiliation{Department of Physics and Research Center OPTIMAS, University of Kaiserslautern, Erwin-Schroedinger-Strasse 46, 67663 Kaiserslautern, Germany}
\author{Patrick M. Buhl}
\affiliation{Johannes-Gutenberg-Universit\"at Mainz, Institut f\"ur Physik, 55128 Mainz, Germany}
\author{Sina Mousavion}
\affiliation{Department of Physics and Research Center OPTIMAS, University of Kaiserslautern, Erwin-Schroedinger-Strasse 46, 67663 Kaiserslautern, Germany}
\author{Bertrand Dup\'{e}}
\affiliation{Nanomat/Q-mat/CESAM,Universit\'e de Li\`ege,B-4000 Sart Tilman, Belgium}
\author{Eva S. Walther}
\affiliation{Department of Physics and Research Center OPTIMAS, University of Kaiserslautern, Erwin-Schroedinger-Strasse 46, 67663 Kaiserslautern, Germany}
\author{Katerina Medjanik}
\affiliation{Johannes-Gutenberg-Universit\"at Mainz, Institut f\"ur Physik, 55128 Mainz, Germany}
\author{Dmitry Vasilyev}
\affiliation{Johannes-Gutenberg-Universit\"at Mainz, Institut f\"ur Physik, 55128 Mainz, Germany}
\author{Sergey Babenkov}
\affiliation{Johannes-Gutenberg-Universit\"at Mainz, Institut f\"ur Physik, 55128 Mainz, Germany}
\author{Martin Ellguth}
\affiliation{Johannes-Gutenberg-Universit\"at Mainz, Institut f\"ur Physik, 55128 Mainz, Germany}
\author{Mahalingam Maniraj}
\affiliation{Department of Physics and Research Center OPTIMAS, University of Kaiserslautern, Erwin-Schroedinger-Strasse 46, 67663 Kaiserslautern, Germany}
\author{Jairo Sinova}
\affiliation{Johannes-Gutenberg-Universit\"at Mainz, Institut f\"ur Physik, 55128 Mainz, Germany}
\author{Gerd Sch\"onhense}
\affiliation{Johannes-Gutenberg-Universit\"at Mainz, Institut f\"ur Physik, 55128 Mainz, Germany}
\affiliation{Graduate School of Excellence Materials Science in Mainz}
\author{Hans-Joachim Elmers}
\affiliation{Johannes-Gutenberg-Universit\"at Mainz, Institut f\"ur Physik, 55128 Mainz, Germany}
\affiliation{Graduate School of Excellence Materials Science in Mainz}
\author{Martin Aeschlimann}
\affiliation{Department of Physics and Research Center OPTIMAS, University of Kaiserslautern, Erwin-Schroedinger-Strasse 46, 67663 Kaiserslautern, Germany}
\author{Benjamin Stadtm\"uller}
\email[]{bstadtmueller@physik.uni-kl.de}
\affiliation{Department of Physics and Research Center OPTIMAS, University of Kaiserslautern, Erwin-Schroedinger-Strasse 46, 67663 Kaiserslautern, Germany}
\affiliation{Graduate School of Excellence Materials Science in Mainz}

%\date{\today}

\pacs{...}
\maketitle
\section{Sample preparation}
\begin{figure}[htbp]
\centering
\includegraphics[width=10.0cm]{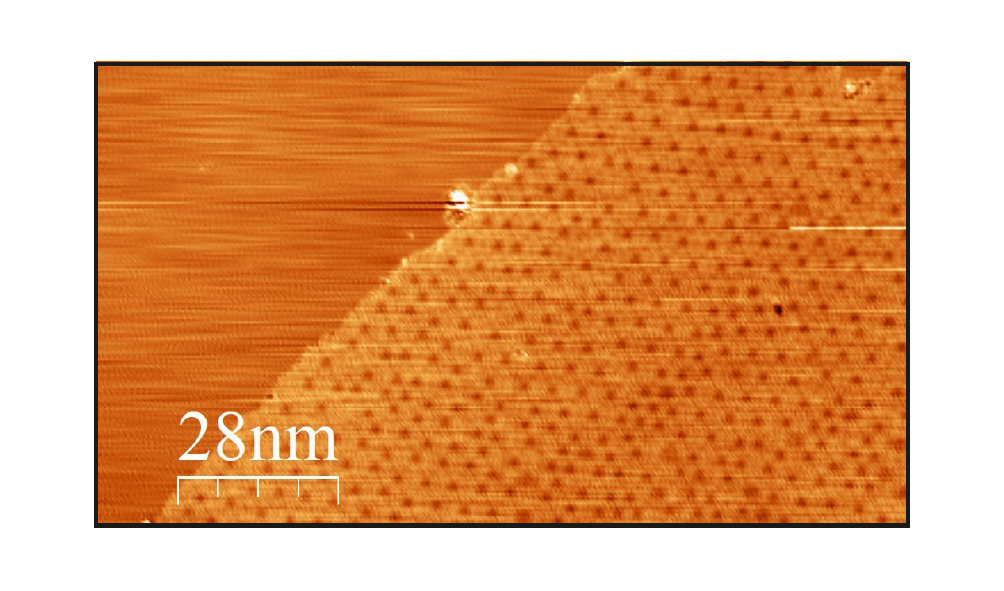}
\caption{Large scale STM image of the DyAg$_2$ surface alloy recorded at room temperature. }
\label{fig:STM}
\end{figure}

The Ag(111) single crystal was cleaned by repeated cycles of argon ion sputtering (E$_{\mathrm{kin}}=1.0-2.0\,$keV, I$_{\mathrm{drain}}\approx 1.5\,\mu$A) followed by sample annealing at T$=730\,$K for 30 minutes. The cleanness of the crystal surface was confirmed by the appearance and narrow line profile of the characteristic LEED spots. The DyAg$_2$ surface alloy was obtained by Dy deposition using a commercial Focus EFM-3 evaporator with a tungsten crucible. During the Dy deposition process, the sample was kept at an elevated temperature of $570\,$K. The Dy coverage was identified qualitatively by  the intensity and linewidth of the diffraction spots of the Dy superstructure and moir\'e spots in LEED. The average deposition time to form a surface alloy is 15 minutes, resulting in a rate of one monolayer in 45 minutes. After deposition, the sample was annealed at the deposition temperature for 15 minutes to desorb non-alloyed Dy and other contamination species. The high sample quality is further confirmed by the homogeneity of the large scale STM image in Fig.~\ref{fig:STM}
\newpage

\section{Simulation of surface umklapp processes}
\begin{figure}[htbp]
\centering
\includegraphics[width=17cm]{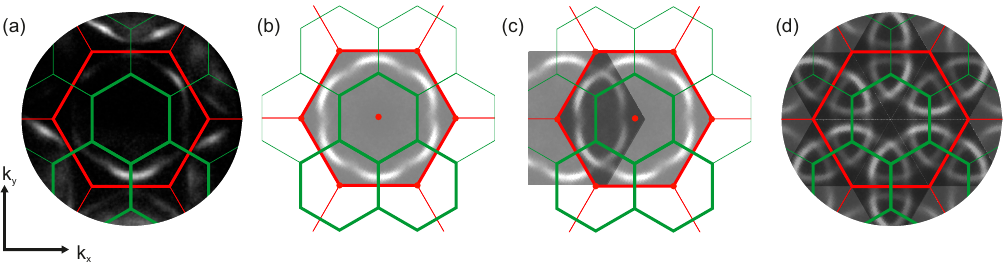}
\caption{Illustration of the individual steps of our procedure to simulate the surface umklapp effect of Ag substrate bands at the DyAg$_2$ superstructure.}
\label{fig:Fig2}
\end{figure}
In this section, we discuss our procedure to simulate the surface umklapp processes of the Ag substrate bands at the DyAg$_2$ superstructure. These simulations are based on experimental data of the clean Ag(111) single crystal surface which were obtained using the identical experimental conditions, i.e., the same momentum microscope as well as the identical light source (identical photon energy and light polarization). 
In the first step, we symmetrized the raw data shown in Fig.~\ref{fig:Fig2}(a) according to the substrate symmetry and eliminate the photoemission intensity of the symmetrized constant energy (CE) maps outside the first surface Brillouin zone (SBZ) of Ag(111), see Fig.~\ref{fig:Fig2}(b). In the next step, we simulated the changes of the photoemission intensity due to scattering of the photoelectrons of the Ag substrate at the periodic superstructure of the DyAg$_2$ surface alloy. During this scattering process, the momentum of the photoelectrons can change by integer multiples of the reciprocal lattice vectors of the DyAg$_2$ superstructure. The latter connect the $\Gamma$-points of the reciprocal lattice of the DyAg$_2$ superstructure which is included in Fig.~\ref{fig:Fig2} as green hexagons. According, we duplicate the emission pattern of the first SBZ of Ag(111) at each $\Gamma$-point of the DyAg$_2$ SBZ. We show this procedure for a single surface umklapp process (single scattering vector) in Fig.~\ref{fig:Fig2}(c). We repeat this procedure for all neighbouring $\Gamma$-points of the DyAg$_2$ SBZ which finally results in the CE map shown in Fig.~\ref{fig:Fig2}(d).
\newpage

\section{Symmetrization of the constant energy maps}

\begin{figure}[htbp]
\centering
\includegraphics[width=17cm]{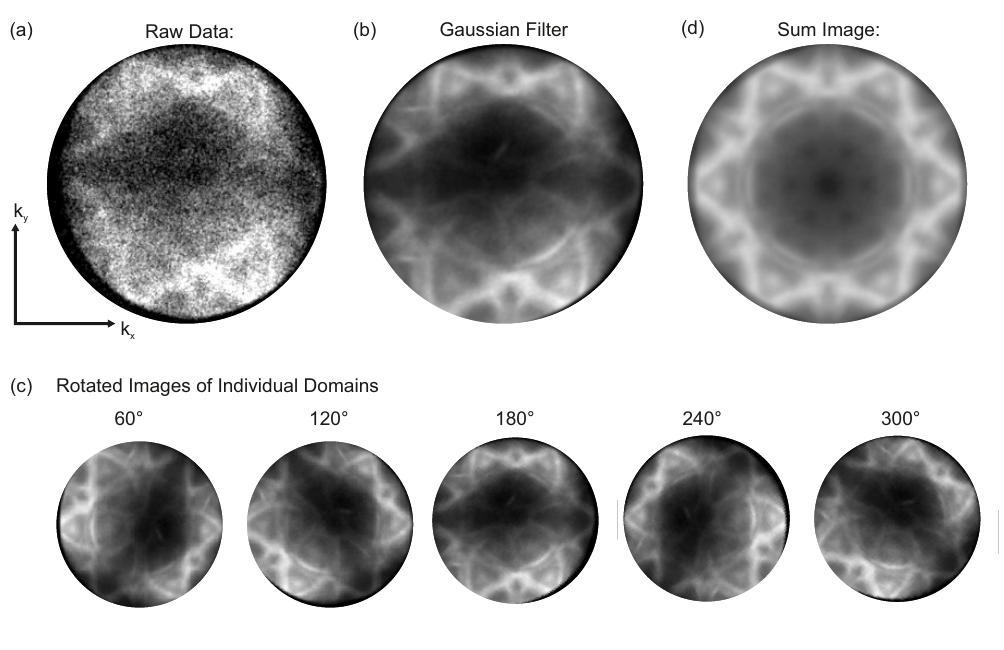}
\caption{Schematic illustration of the data symmetrization procedure to compensate for artificial intensity asymmetries due to the experimental geometry.}
\label{fig:Fig1}
\end{figure}

The CE maps presented in the manuscript were symmetrized according to the symmetry of the DyAg$_2$ surface reconstruction to compensate for artificial intensity asymmetries due to the experimental geometry. The most obvious influence of the experimental geometry (grazing incidence of light) and the light polarization (horizontal- or s-polarized synchrotron radiation) is the horizontal black stripe with almost vanishing intensity in all CE maps, see for instance in Fig.~\ref{fig:Fig1}(a) and (b). 
The data treatment procedure is shown for an energy slice with a binding energy of E$_\mathrm{B}=0.8\,$eV. The corresponding CE map of the raw data is shown in Fig.~\ref{fig:Fig1}(a). In the first step, we applied a Gaussian 3D filter implemented in the software package \textit{ImageJ} to improve the image quality of the entire 3D momentum microscopy stack for all energy slices recorded in a single experiment. The change of the data quality is illustrate for the energy slide at E$_\mathrm{B}=0.8\, $eV in panel (b). Although the Gaussian 3D filter reduces the spectroscopic resolution of the 3D data cube, we do not suppress any spectroscopic information due to the intrinsically large linewidth of the Dy-Ag bands in energy and momentum space. 

In the next step, we considered the six fold symmetry of the surface alloy superstructure. Therefore, the CE map in Fig.~\ref{fig:Fig1}(b) was rotated around the $\Gamma$-point by $n\times 60^{\circ}$, n$=0,1,2,4,5$. The six rotated CE maps are shown in Fig.~\ref{fig:Fig1}(c). Finally, the symmetrized CE maps are obtained by summing up these 6 CE maps corresponding to the different domains of the DyAg$_2$ surface alloy. The CE maps after this symmetrization procedure are shown in the main manuscript (Fig.~2(b) ) and in Fig.~\ref{fig:Fig1}(d) of the supporting information. 

This data treatment procedure does not only allow us to improve the image quality of the CE maps, but also to suppress minor distortions of the CE maps due to aberrations of the momentum microscopy optics. The uncertainties of this data treatment procedure (not ideal identification of $\bar{\Gamma}$-point, instrumental distortions,\dots) result in a minor broadening of the spectroscopic features in momentum space. 
This data treatment procedure was simultaneously applied to all CE maps of the entire 3D data stack. 
\newpage

\section{Valence Band Spectroscopy of the Dy 4f states of the DyAg$_2$ surface alloy}
\begin{figure}
\centering
\includegraphics[width=17cm]{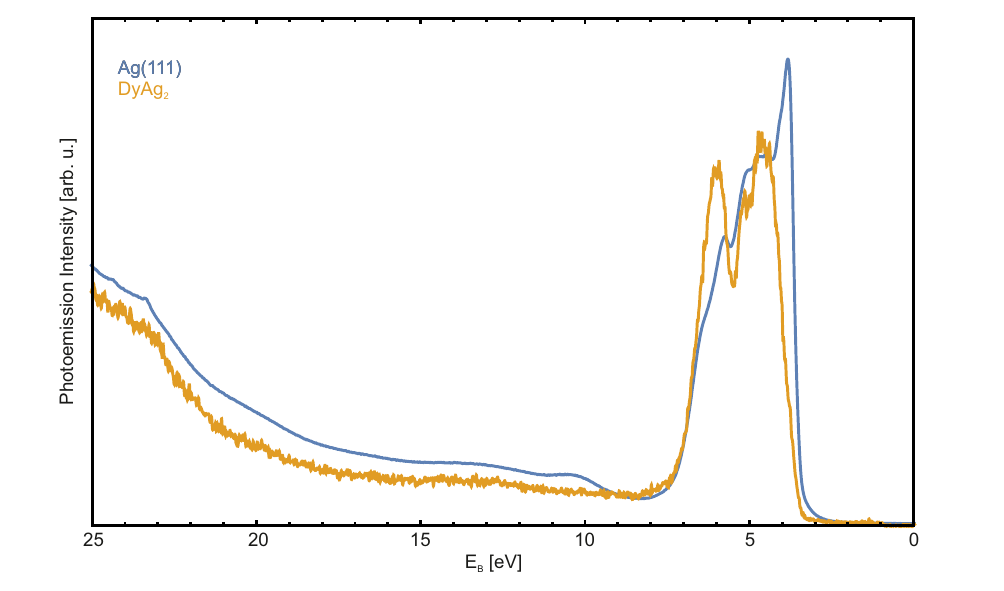}
\label{fig:4f}
\caption{Momentum integrated photoemission yield of the extended DyAg$_2$ valence band structure recorded with a hemispherical electron analyzer and He-II radiation at $\bar{K}$-point of DyAg$_2$ surface Brillouin zone.}
\end{figure}

In the main manuscript, we report exceptionally small binding energies for the spin split Dy 4f states of E$_\mathrm{B}=0.2\,$eV and E$_\mathrm{B}=1.8\,$eV, respectively. Such small binding energies of the Dy 4f states are in strong contrast to photoemission studies of Dy bulk crystals and Dy thin films for which the Dy 4f states are typically observed at binding energies larger than $4\,$eV. For a Dy thin film, Lang et al. reported the formation a Dy multiplet with at least five characteristic peaks in the binding energy range between $4\,$eV and $12\,$eV \cite{Lang1981}. These large binding energies are characteristic for elemental lanthanide materials and are attributed to the strong electronic correlations of the localized 4f electrons in conjunction with final state effects in the photoemission process. 

In the light of these previous findings, we recorded additional photoemission data in a larger binding energy range from E$_\mathrm{B}=0\,$eV and E$_\mathrm{B}=25\,$eV to support our assignment of the Dy4f states to spectroscopic features in direct vicinity of the Fermi energy. The corresponding photoemission data of the DyAg$_2$ surface alloy are shown in Fig.~4 as orange curve. As a reference, we recorded the same data for a bare Ag(111) surface (blue curve in Fig.~4). These data were recorded with a hemispherical electron analyser and monochromatic He-II$_\alpha$ radiation. 

The most obvious difference between both spectra is a change of the spectral density in the energy range of the Ag 3d bands between $3\,$eV and $7\,$eV. Similar changes are frequently observed for adsorbate structures on metal surfaces and can be attributed to photoelectron diffraction of substrate d photoelectrons at ordered or disordered superstructures \cite{Bocquet.2011,Stockl.2018}.

Most importantly, the photoemission yield of the DyAg$_2$ surface alloy is featureless for binding energies larger than $7\,$eV. This clearly shows that the binding energy of the Dy 4f states is significantly altered by the formation of a DyAg$_2$ surface alloy. Hence, we can unambiguously attribute the spectroscopic features at E$_\mathrm{B}=0.2\,$eV and E$_\mathrm{B}=1.8\,$eV to the Dy 4f states of the Dy alloy atoms. 

Both localized Dy 4f states reveal a characteristic emission pattern in momentum space which are shown in the CE maps in Fig.~5a,b. These CE maps were recorded at the NanoEsca endstation at the synchrotron Elettra, Trieste, using a photon energy of $50\,$eV and p-polarized synchrotron radiation. The red hexagons in both CE maps mark the surface Brillouin zone of the Ag(111) surface, the green hexagons the one of the DyAg$_2$ surface alloy. Most importantly, only the CE map at E$_\mathrm{B}=0.2\,$eV reveals a significant spectral density at the $\bar{K}$-point of the Ag(111) surface Brillouin zone. The latter coincides with the $\bar{\Gamma}$-point of the second surface Brillouin zone of the DyAg$_2$ surface alloy. In contrast, the CE map at E$_\mathrm{B}=1.8\,$eV shows the largest spectral density at the $\bar{M}$-point of the Ag(111) surface Brillouin zone. This is even more clearly visible in the partial yield spectra in Fig.~5c which have been extracted at selected high symmetry points in momentum space, which are marked by coloured points in the CE map in Fig.~5a. These supporting data demonstrate the characteristic momentum space signatures of the localized Dy 4f states. It also illustrates that only the Dy 4f state at E$_\mathrm{B}=0.2\,$eV exhibits spectral intensity in the energy vs. momentum cut along the $\bar{\Gamma}$-$\bar{M}$ direction of the DyAg$_2$ surface Brillouin zone discussed in Fig.~2d of the main manuscript. 

\begin{figure}
\centering
\includegraphics[width=17cm]{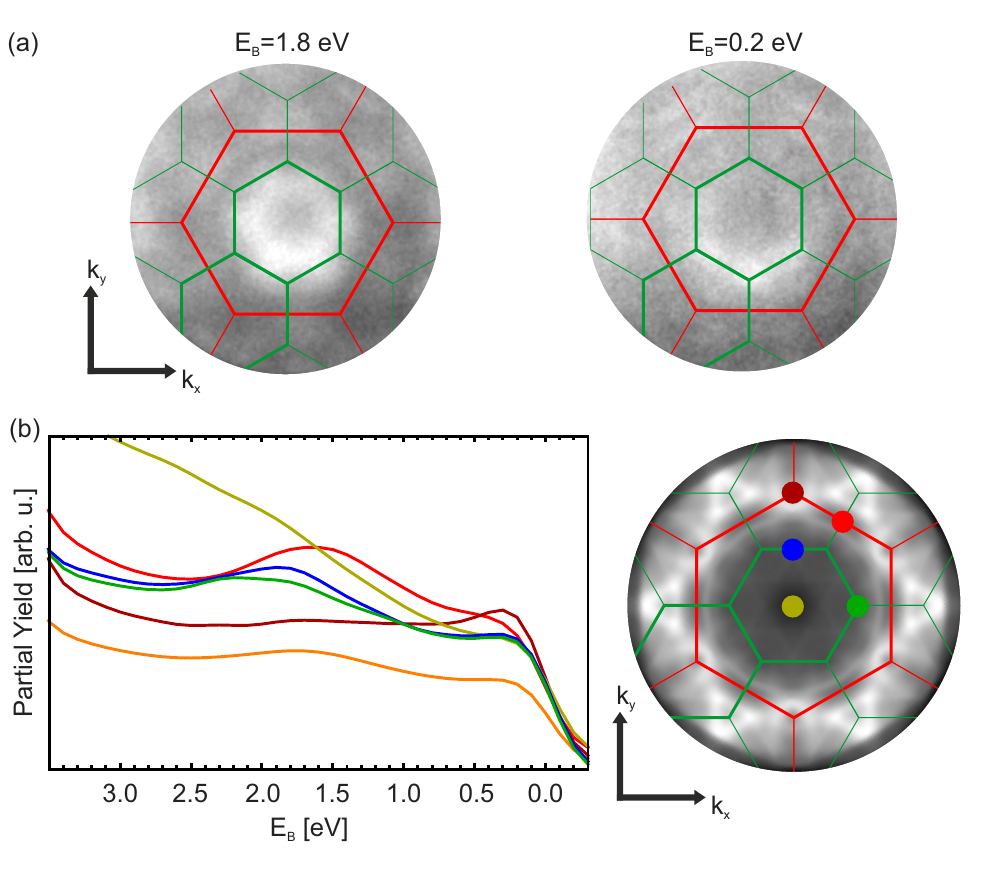}
\label{fig:4f_k}
\caption{(a) CE maps of the Dy 4f majority and minority momentum pattern obtained with p-polarized light, grazing incidence of light and a photon energy of h$\nu=50\,$eV. (b) Partial yield spectra extracted at selected positions in momentum space, which are marked in the CE maps shown in the right part of the panel. }
\end{figure}
\newpage

\section{sample magnetization and spin-resolved momentum microscopy}
\begin{figure}[htbp]
\centering
\includegraphics[width=17cm]{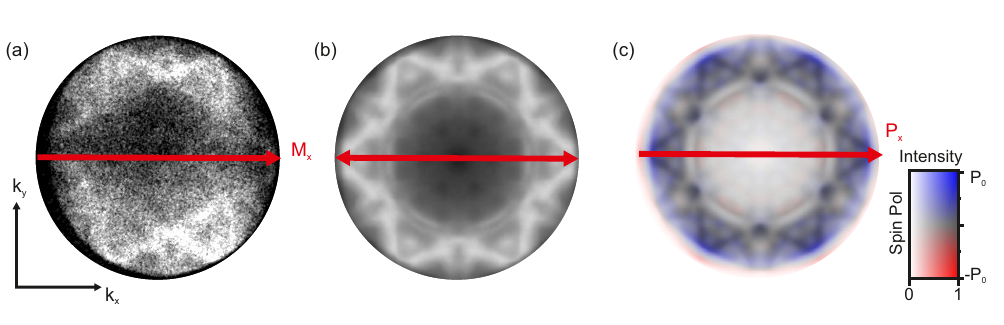}
\caption{Magnetization direction and spin sensitivity axis of the imaging spin filter used in our spin-resolved momentum microscopy experiment.}
\label{fig:Fig3}
\end{figure}

In this work we use an imaging spin filter with an Ir(100) scattering target \cite{Kolbe2011,Kutnyakhov2013}. 
The sample was magnetized in the plane of the sample surface along the horizontal "x"-direction (as indicated in Fig.~\ref{fig:Fig3}) which corresponds to the $\bar{\Gamma}-\bar{M}$ direction of the Ag(111) surface Brillouin zone and to the $\bar{\Gamma}-\bar{K}$- direction of the DyAg$_2$ surface alloy. This direction also coincides with the sensitivity axis of our spin filter. 
A detailed characterization of this type of spin filter was recently reported by Kutnyakhov et al.\cite{Kutnyakhov2013}. The spin asymmetry is directly correlated to the LEED-IV curve of the Ir(100) scattering target and hence dependents crucially on the scattering energy. Spin contrast can be obtained by recording the photoelectron pattern after the scattering target for two characteristic scattering energy with different spin-dependent cross sections. In this work, we used the scattering energies E$_S$=12.5 eV and E$_s$=19.0 eV. These scattering energies have a spin asymmetry of -18\% for 12.5 eV and $\approx$-3\% for 19.0 eV. These values lead to a spin sensitivity of our spin filter of 15\% .
The quantitative analysis is discussed in detail by C. Tusche et al.\cite{Tusche2013}. At this point we only want to point out that the Sherman function and the reflectivity of the imaging spin filter do not only depend on the scattering energy but also on the polar and azimuth angle as well. These have been taken into account in our data analysis procedure. The latter was performed using a dedicated data analysis script developed by C. Tusche et al. \cite{Tusche2013}.
\newpage

\section{Density functional theory calculations}

Here, we provide additional information regarding the density functional theory (DFT) calculations. 
As pointed out in the manuscript, we conducted our DFT calculations for a large variety of Hubbard parameters $U$ between $0\,$eV and $7\,$eV. The band structure as well as the corresponding spin-dependent density of states of the Dy 4f electrons are shown in Fig.~7a,b for different Hubbard parameters U. The most significant change of the band structure is an energy shift of the Dy 4f states depending on U. Increasing U from $0\,$eV to $4\,$eV results in a shift of the Dy 4f minority and majority states to larger binding energies. A further increase of U leads again to a decrease of the binding energy of the Dy 4f states. However, the Dy 4f states are located within the binding energy interval between $0\,$eV and $2\,$eV for the entire parameter set of U. The best agreement between experiment and theory was obtained for $U=1.5\,$eV.

\begin{figure}[bp]
\includegraphics[width=17cm]{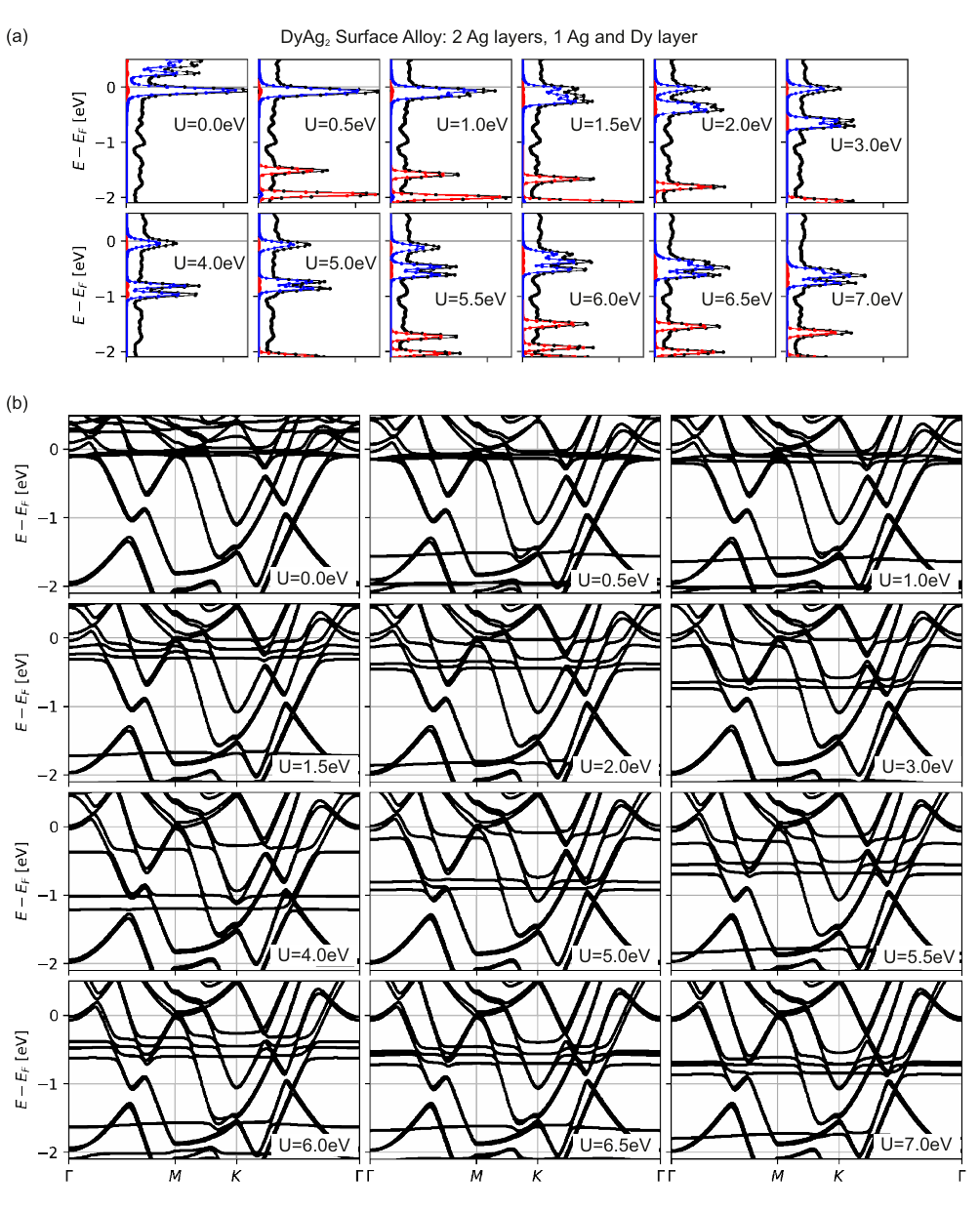}
\caption{DFT+U calculations for DyAg$_2$ surface alloy on a slab of 2 layers of Ag for different U. (a) Total DOS of 4f states of Dy. (b) band resolved DFT.}
\label{fig:DFT_U}
\end{figure}

\begin{figure}
\includegraphics[width=17 cm]{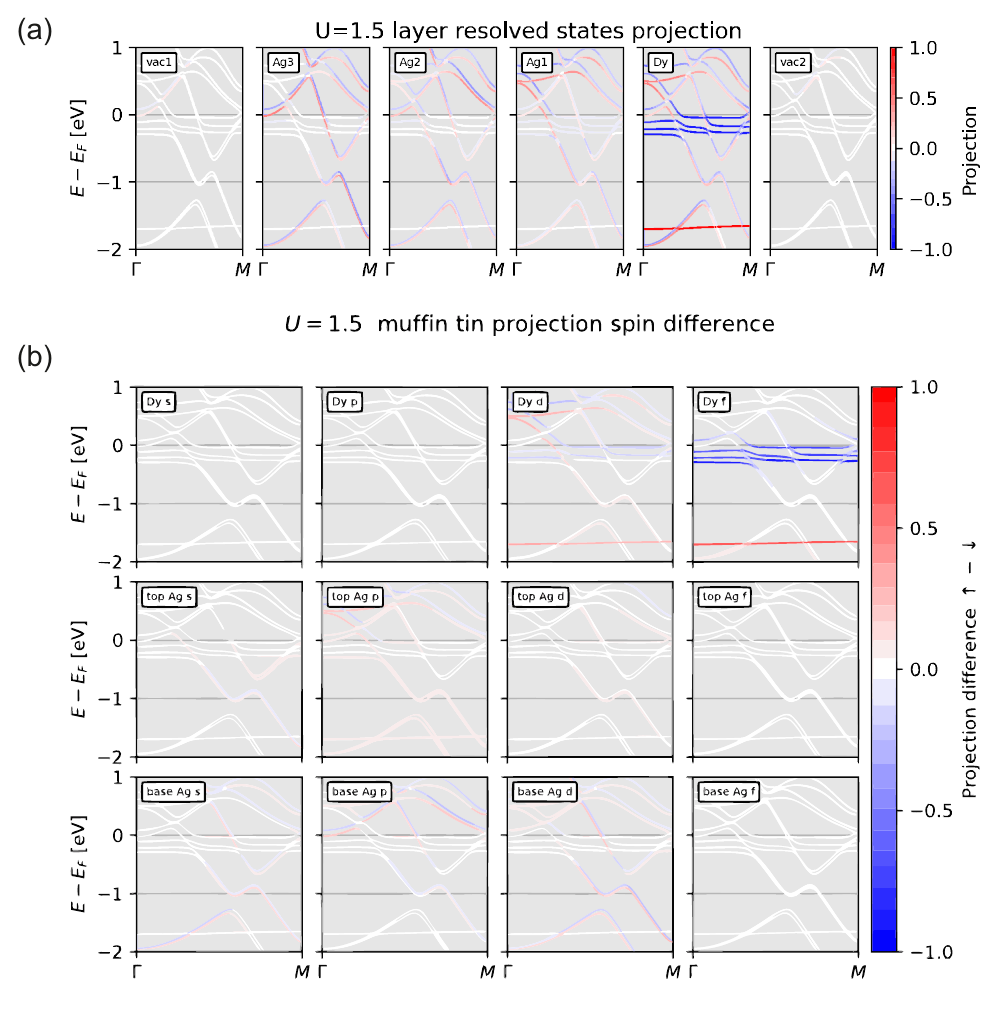}
\caption{DFT+U calculations for U=1.5 for DyAg$_2$ on a slap of 3 layers of Ag. (a) projection of the states on the different layers. Ag1 represents the alloy layer. (b) projection of the states onto the different Dy states (top row), Ag surface states (middle row) and Ag base states (bottom row).}
\label{fig:projections}
\end{figure}

More information about the orbital character and locations of the different electronic states can be obtained by projection of the states on the different layers and onto the muffin tin spheres of the atomic orbitals. We first want to illustrate the layer dependent band structure. We show the calculations for U=1.5 in Fig.~\ref{fig:projections}(a). The 4f-derived states are only projected on the Dy layer. The Dy-Ag hybrid surface state at the $\bar{\Gamma}$-point is located in the Dy layer as well as in the topmost Ag layer. The backfolded Ag sp-bands are mainly located in the lower layer of the Ag slap. The parabolic state with an onset at E$_\mathrm{B}=2\,$eV is a quantum well state due to the finite vertical size of the Ag slab and shift to larger binding energies when increasing the vertical size of the Ag slab. It is a result of the slap calculation and can therefore not be observed in the experiment. \\
The orbital character of the states is illustrated in the orbital projects band structure in Fig.~\ref{fig:projections}(b). The top row shows the band structure projection onto the Dy orbitals. We observe only a vanishing s and p character of the bands of the entire valence band structure of the surface alloy. Most importantly, the Dy-Ag hybrid surface states exhibits mainly Dy d-orbital character with small contributions of Dy 4f states. In addition, the Dy-Ag hybrid surface consists mostly of Ag p-orbitals from atoms of the topmost Ag layer (middle row). The bulk-like Ag layer of the slap (bottom row) contribute largely with s-,p-, and d-like orbitals to the valence band structure via back folded substrate bands. In this way, we can conclude that the Dy-Ag hybrid surface state is mainly the result of the hybridization between Ag p-and Dy d-states. 

\newpage
\newpage

\section{Band Structure Calculations vs. Band Structure Cut through ARPES Data}
\begin{figure}
\includegraphics[width=17 cm]{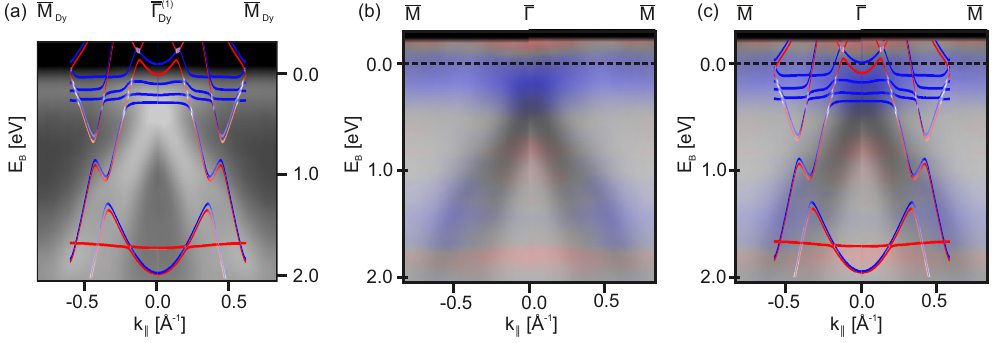}
\caption{(a) Spin integrated energy vs. momentum cut through our momentum microscopy data of the DyAg$_2$ surface alloy. We superimposed these data with the density functional theory calculations showing the relevant bands in this energy and momentum range. (b) Normalized intensity of the spin resolved electronic valence structure of DyAg$_2$ to reveal the exchange split hybrid interface state above E$_F$. (c) Normalized spin resolved electronic valence structure of DyAg$_2$ versus density functional theory calculations showing the relevant bands in this energy and momentum range.}
\label{fig:ARPES_vs_DFT}
\end{figure}

In this section, we superimpose the band structure calculation onto our ARPES data to support the assignment of the different states to spectroscopic features in the main manuscript. We show the spin integrated electronic valence structure observed by momentum microscopy in Fig.~\ref{fig:ARPES_vs_DFT}(a). We observe a very good agreement between the calculated band structure and the measured intensity distributions. In Fig.~\ref{fig:ARPES_vs_DFT}(b), we present the normalized spin resolved electronic valence structure of DyAg$_2$ to reveal the lower branch of the exchange split hybrid interface state of DyAg$_2$. In Fig.\ref{fig:ARPES_vs_DFT}(c) we superimpose the DFT results onto the spectra shown in (b). We observe excellent agreement between the calculations and the measured spin resolved band structure.
